\documentclass[twocolumn, tighten, times]{aastex7}

\begin{document}

\shortauthors{Gordon \& Law}
\shorttitle{MRS PFPCs}

\title{JWST MIRI Medium Resolution Spectrometer Point Fixed Pattern Corrections: \\ Cleaner and Higher Signal-to-Noise Spectra of Point Sources}

\author[0000-0001-5340-6774]{Karl~D.~Gordon}
\affiliation{Space Telescope Science Institute, 3700 San Martin
  Drive, Baltimore, MD, 21218, USA}
\affiliation{Department of Physics and Astronomy, Universiteit Gent, Proeftuinstraat 86 N3, B-9000 Ghent, Belgium}
\email[show]{kgordon@stsci.edu}

\author[0000-0002-9402-186X]{David R.\ Law}
\affiliation{Space Telescope Science Institute, 3700 San Martin Drive, Baltimore, MD, 21218, USA}
\email[hide]{dlaw@stsci.edu}

\submitjournal{AJ}
\accepted{13 Aug 2026}

\begin{abstract}
The JWST Mid-Infrared Instrument Medium Resolution Spectrometer provides the capability to obtain spectra from 5--28~\micron.
The JWST data reduction pipeline removes the majority but not all of the instrument artifacts and the signal-to-noise (S/N) of the resulting spectra are limited by fixed pattern noise.
Building on previous work, Point Fixed Pattern Corrections (PFPCs) are constructed using observations of O, A, and G dwarf flux calibration stars and asteroids taken using the default four point dither pattern.
The PFPCs can be applied to spectra of point sources taken with target acquisition and the same dither pattern.
They can be used alone or with the pipeline residual fringe correction depending on the sources spectral properties.
Both narrow and broad artifacts are removed by the PFPCs improving the spectra regardless of their S/N.
For higher S/N observations, the PFPCs significantly improve the S/N by up to factors of a few and S/N values of 1000 or more.
The MRS-PFPC python package is provided to allow anyone to utilize the PFPCs for their own data.
\end{abstract}

\keywords{}

\section{Introduction}
\label{sec_intro}

The Mid-Infrared Instrument \citep[MIRI,][]{Rieke15, Wright23} Medium Resolution Spectrometer \citep[MRS,][]{Wells15, Argyriou23} provides the capability to observe medium resolution spectroscopy ($R \sim 3000$) from 5--28~\micron\ for the James Webb Space Telescope \citep[JWST,][]{Gardner23}.
The MRS is composed of four nested Integral Field Units (IFUs) designated channels 1, 2, 3, and 4 each having three grating settings A/SHORT, B/MEDIUM, and C/LONG.
The multiplexing is such that with three exposures, spectra covering the full 5--28~\micron\ can be taken.

The JWST data reduction pipeline \citep{Bushouse26} corrects for many instrumental signatures, yet the signal-to-noise (S/N) is limited by fixed pattern noise \citep{Gasman23, Pontoppidan24}.
Fixed pattern noise is often seen in spectroscopic data \citep{Leckrone89, Heap95} and is often corrected by deriving a fixed pattern correction using sources with known spectral shapes \citep{Massa20, Gasman23, Pontoppidan24}.
Absent a correction, the fixed pattern noise fundamentally limits the S/N of MRS spectra and the weakest features that can be detected.
Sources of fixed pattern noise include flat field uncertainties, wavelength dependent flux calibration uncertainties, sampling artifacts, and residual fringes.

For MIRI MRS, residual fringes are known to be the major source of fixed pattern noise.
MRS spectra show strong fringes on order of 10-30\% as the observed wavelengths are similar to the thicknesses of the detector and coatings \citep{Argyriou20}.
The majority of the fringe signal is corrected using a static fringe flat that is derived from observations of extended sources.
The fringing produced by a point source is different than that of an extended source however and varies as a function of the location of the source on the detector; this mismatch results in point source spectra showing residual fringes on the order of 1-2\% amplitude after correction using the static fringe flat alone.
The jwst pipeline optionally corrects these residual fringes by fitting
a series of Fourier components in the known range of fringe frequencies to the spectral data to derive a correction for each MRS spectral segment.
However if the source has regularly-spaced emission or absorption lines with a similar frequency to the residual fringes (e.g., due to molecular absorption bands), then the residual fringe correction can remove astrophysical signal as well as the residual fringes.

While the residual fringe correction in the jwst pipeline focuses on correcting the primary type of fixed pattern noise there are other sources of such noise for MRS spectra as well.
As discussed by \citet{Law25}, the MRS flux calibration in the jwst pipeline is tied to observations averaged across many locations in the IFU field of view.  Most point source observations are obtained using a small number of dither positions however, for which the $\sim$1.5\% flatfield uncertainties in the 5-18 $\micron$ range can limit performance in the high-SNR regime. 
Likewise, some percent-level absorption artifacts are known in MRS spectra \citep[e.g., one at 5.8 $\micron$ feature described by][]{Decleir25, Zeegers25} that are likely instrumental in origin but which are too broad to be due to fringing.

Motivated by observations of point sources with spectra too complicated for the jwst pipeline residual fringe correction step, MRS fixed pattern corrections for such sources have been derived using observations of an A star \citep{Gasman23} and two asteroids/one A star \citep{Pontoppidan24}.
The fixed pattern corrections were measured for each of the standard four dither positions as the fixed pattern noise will be different as the location on the detector changes.
In both cases, the fixed pattern correction was measured from the ratio of the source spectrum reduced without residual fringe correction and a model of the star or asteroids.
Being based on an A star, the \citet{Gasman23} correction focused on the MRS channels 1-3 as stars are bright at shorter MRS wavelengths.
While the \citet{Pontoppidan24} correction focused on MRS channels 2-4 using asteroids that are bright at the longer MRS wavelengths, it did use an A star for channel 1 providing corrections for all MRS wavelengths.
These corrections provided significant improvement over the static fringe flat alone in the standard jwst pipeline reductions, delivering expected S/N values at least 300 and significantly higher S/N values up to 700 in selected wavelength ranges.

Similarly, driven by variations in the fixed pattern noise due to small, sub-pixel variations in the actual pointings for each dither position \citep[largely due to non-repeatability in the MRS grating wheels; see discussion by][]{Patapis24}, \citet{Gasman24, Gasman25} obtained measurements of the flux calibration O9V star 10 Lac in intra-pixel dithers around each standard MRS dither position.
These observations allow for fixed pattern corrections that are specific to the delivered sub-pixel position at each dither position, a further improvement over a single correction for each dither position.
The performance of these corrections produces significant improvements for the spectra for individual pixels, but only a modest improvement over the dither averaged spectra.
Specifically, the improvements are largest in MRS channel 2 resulting in spectra with $\sim$0.5\% fringes.
For the other MRS channels the residuals are $\sim$1\% for this correction, similar to that seen with the default jwst pipeline corrections.

In this contribution, we build upon the existing dither dependent fixed pattern corrections of \citet{Gasman23} and \citet{Pontoppidan24}.
Our main goal is to combine the results from both stars and asteroids to provide the highest S/N fixed pattern correction for the default four MRS dither positions, pushing the calibration of MRS point source spectra into the sub-1\% regime.
Combining multiple stars and asteroids not only provides for higher S/N, but it also allows for regions with absorption or emission lines in the stellar spectra to be masked and filled in with the corrections derived from the asteroid spectra.
This work focuses on deriving a single, wavelength dependent fixed pattern correction for each dither position for each MRS channel and grating position.
It does not attempt to derive corrections that account for the delivered sub-pixel locations at each dither position using the \citet{Gasman23, Gasman24} observations.
Given the excellent repeatability of the JWST pointing when combined with Target Acquisition \citep{Rigby23}, it is not unreasonable to expect that a single correction for each dither position may provide a good enough correction without accounting for small shifts around the nominal dither positions.

The observations and data reduction of the stars and asteroids are described in \S\ref{sec_data}.
The construction of the Point Fixed Pattern Corrections (PFPCs) are detailed in \S\ref{sec_pfpc} with their properties and quantitative relation to residual fringes.
How to use the PFPCs using the MRS-PFPC package\footnote{\url{https://github.com/STScI-MIRI/MRS-PFPC}}, the spectral artifacts removed, and improvement in S/N is elaborated in \S\ref{sec_results}.
The summary is provided in \S\ref{sec_summary}.

\section{Data}
\label{sec_data}

\begin{deluxetable}{lcc}
\label{tab_targets}
\tablecaption{Calibration Targets}
\tablehead{\colhead{Target} & \colhead{SpT} & \colhead{PID}}
\startdata
10 Lac & O9V & 3779 \\
$\mu$ Col & O9.5V & 4497 \\
$\delta$ UMi & A1Van & 1536 \\
HR 5467 & A1V & 4496 \\
HD 2811 & A3V & 1536, 4496, 6604 \\
HR 6538 & G1V & 4498 \\
HD 37962 & G2V & 1538 \\
16 Cyg B & G3V & 1538 \\
HD 167060 & G3V & 1538 \\
515 Athalia & asteroid & 1549 \\
526 Jena & asteroid & 1549 \\
\enddata
\end{deluxetable}

The observations of point sources used to construct the PFPCs are listed in Table~\ref{tab_targets}. 
All the observations were taken with the default dither pattern with the parameters 4-POINT, NEGATIVE, POINT-SOURCE, and ALL\_MRS.
They are the combination of MRS flux calibration observations taken in cycles 1--3 \citep{Gordon22, Law25} and two asteroids \citep{Pontoppidan24}.
The flux calibration observations targeted stars stars with O, A, and G spectral types providing spectra with absorption lines of varying strengths.
The asteroids observed are in the Themis family, are point sources at MIRI wavelengths, and have effectively featureless mid-IR spectra.  Since many of these asteroid are time-variable they are not ideal for
absolute flux calibration or relative calibration between spectral bands, but can provide excellent constraints on the spectral response
within a given band due to their nearly ideal blackbody shapes.
Stars provide excellent absolute and relative flux calibration \citep{Law25}, except for the regions near strong lines.
For the PFPC, the goal is to measure the fixed pattern noise relative to the existing flux calibration.
The combination of asteroids and stars provides high S/N measurements at short and long MIRI wavelengths.

To prepare the data for measuring the PFPC, the observations were reduced with jwst pipeline version 2.0.0 and the following specific configurations:
\begin{enumerate}
\item The detector1 stage is run with default parameters.
\item For the spec2 stage, 
\begin{enumerate}
\item the bad pixel self calibration step (badpix\_selfcal) was run to improve the detection of bad pixels,
\item the pixel\_replace step was run with the mingrad algorithm,
\item the cube building was done by band and in IFU coordinates (output\_type = band and coord\_system = ifualign), and 
\item the autocentroiding was enabled for the extract\_1d step (ifu\_autocen option).
\end{enumerate}
\item The spectral leak seen in the 3A extracted spectrum was corrected using the 1B segment spectrum.
\end{enumerate}
Steps 1--3 are done automatically using the MRS-PFPC \texttt{pfpc\_proc} command.
The result of this reduction is a 1d extracted spectrum at each of the 4 dither positions for each of the 12 combinations of channel (1, 2, 3, 4) and grating (A, B, C).
The spectral extractions were done using the default parameters that match those used for the pipeline flux calibration \citep{Law25}.
The background is subtracted using the default annular aperture and is small compared to the source flux for stars in channels 1--3 and asteroids for channels 2--4.
For the spectral cube construction, the Adaptive Trace Modeling option \citep{Law26} was not used for this work as the apertures used are large enough not to suffer from resampling artifacts (except for one specific case discussed in the next section).

\section{Point Fixed Pattern Corrections}
\label{sec_pfpc}

\begin{figure*}[tbp]
\epsscale{1.1}
\plottwo{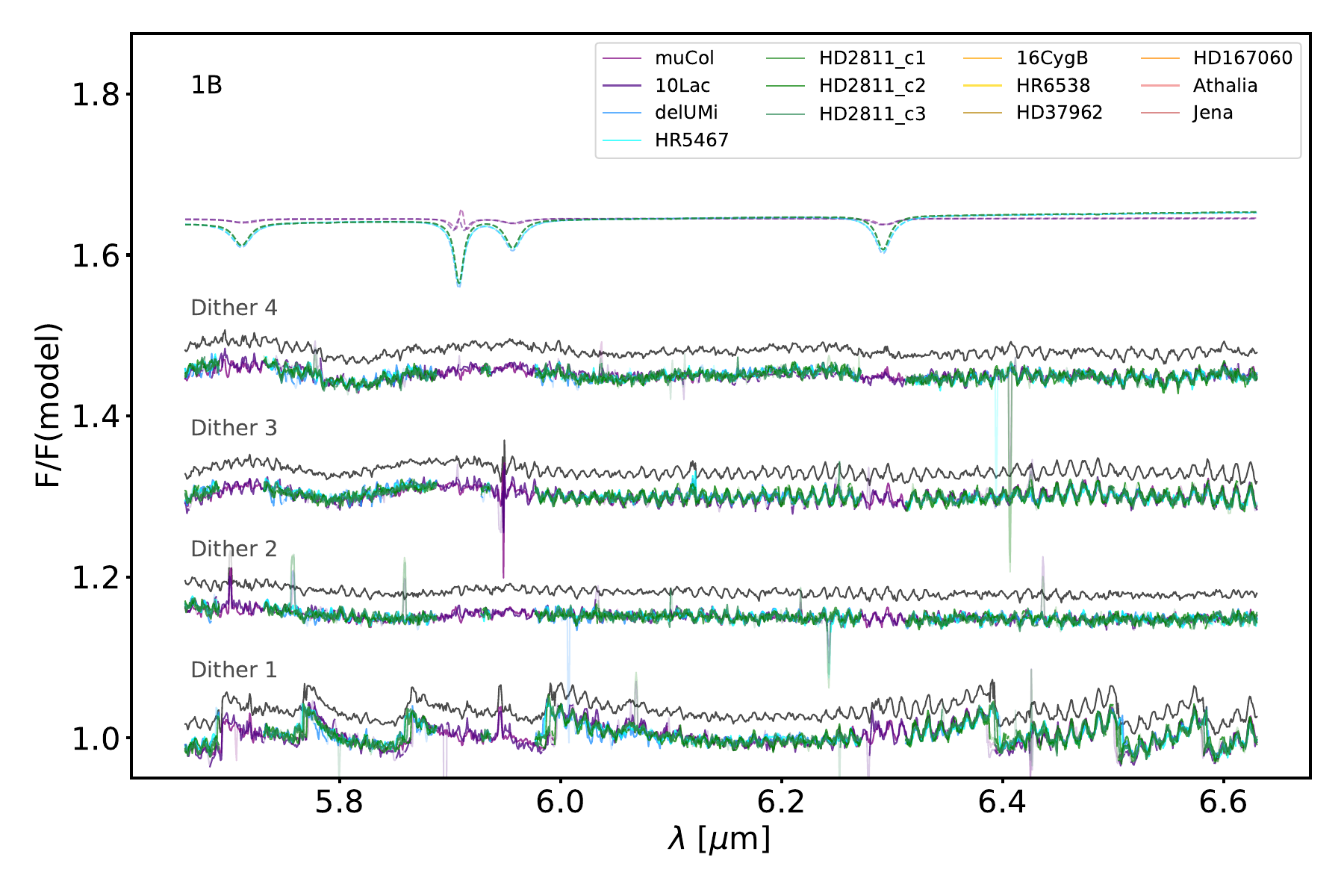}{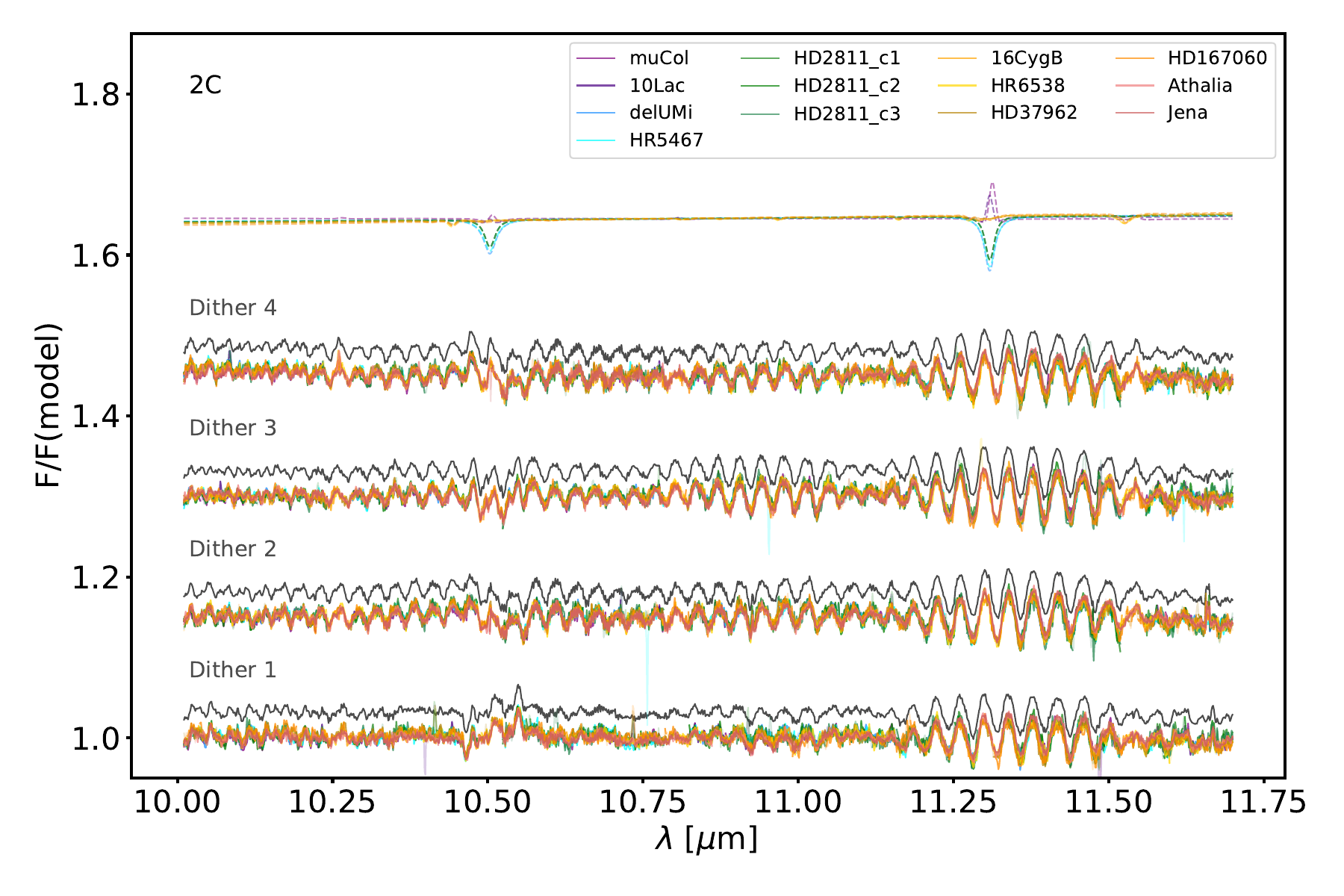} \\
\plottwo{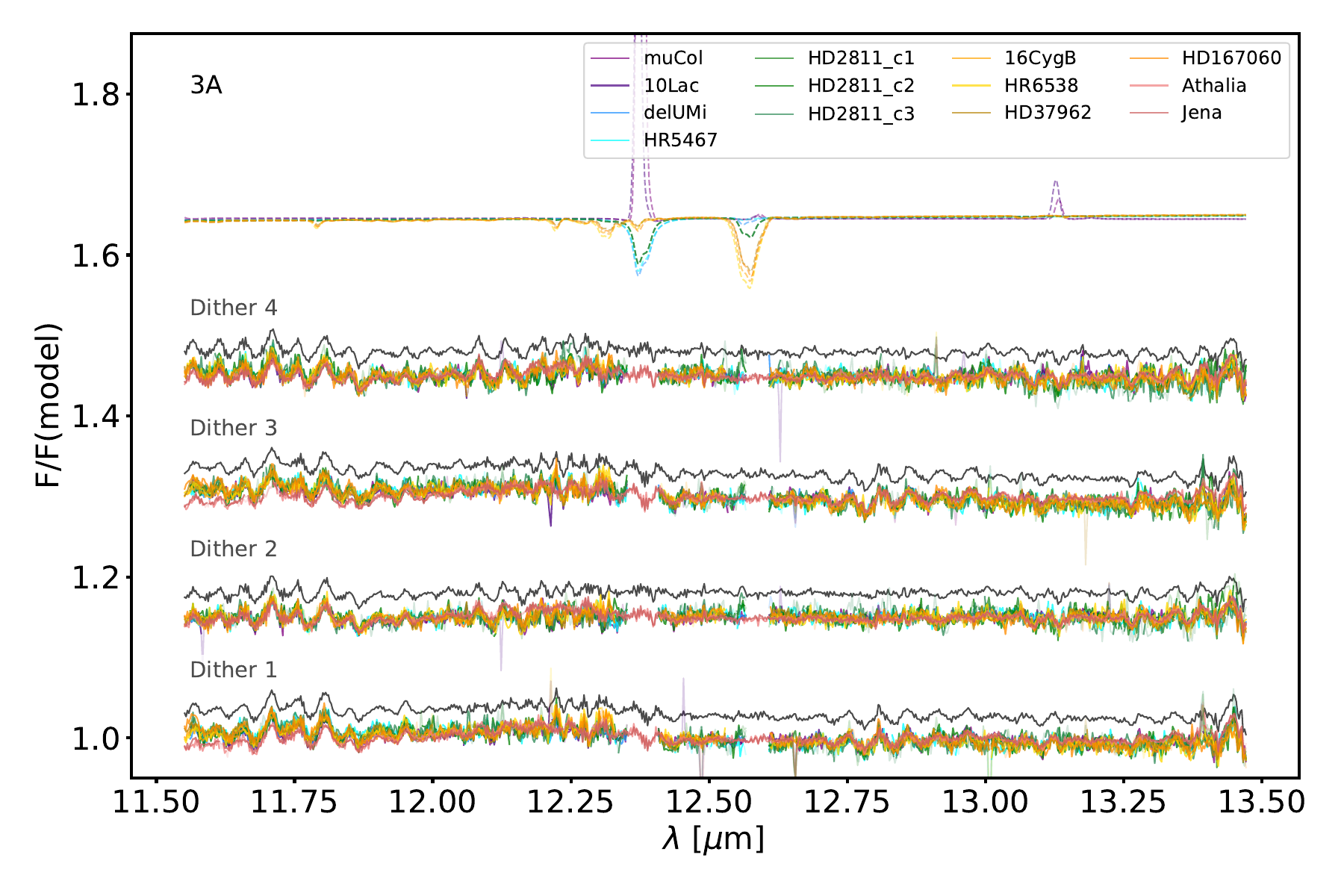}{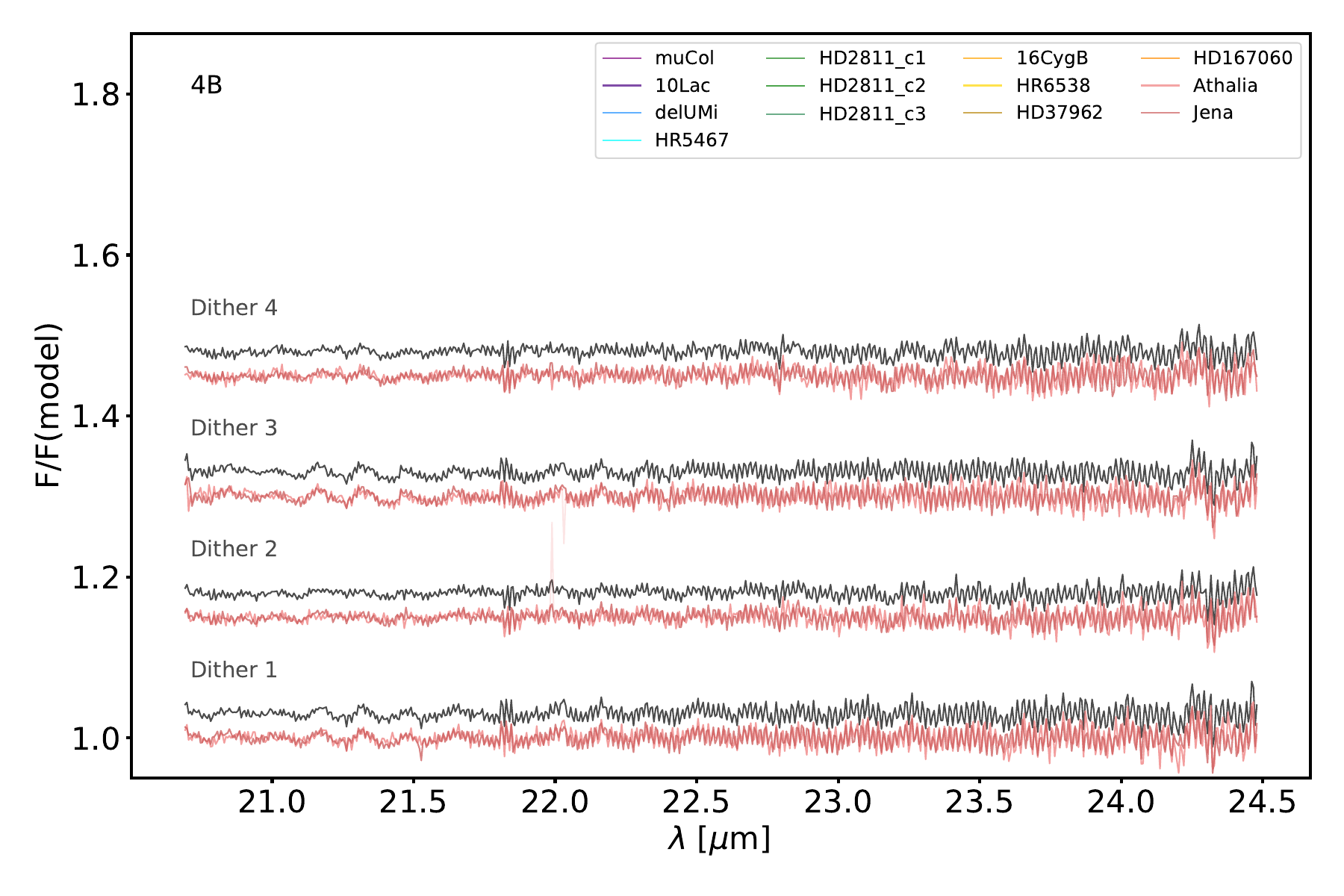}
\caption{The individual PFPCs for channels/gratings 1B, 2C, 3A, and 4B for each star and asteroid are plotted for each dither position offset for clarity.
The sigma-clipped average PFPC for each dither is plotted in black offset above the respective individual PFPCs.
The clipped data is shown fainter than the data used to construct the averages.
The models used for each star are shown at the top of each panel.
The models for asteroids Jena and Athalia are quadratic fitf to each PFPC and are not plotted.
The asteroids are not used for channel 1 and the stars are not used for channel 4.
Regions near strong lines for the stars that may be poorly corrected by the models are not plotted nor used for the averages.
\label{fig_pfpc_stability}}
\end{figure*}

The PFPCs are measured for each of the 12 MRS spectral segments and 4 dither positions independently resulting in 48 wavelength dependent corrections.
A measurement of the PFPC can be done for each spectral segment by dividing the observed spectrum by the appropriate model.
As the stars are all flux calibration standards, the CALSPEC \citep{Bohlin14} models are used for them.
For the asteroids, the model is a quadratic fit to the spectrum in each spectral segment as while these asteroids can show broad weak features \citep{Licardro12}, a quadratic does a good job of removing the continuum and any such broad features.
This was confirmed by the excellent agreement in the PFPC measurements between the asteroids and the stars.
Depending on the source, there are regions near absorption lines, emission lines, or simply have low S/N that are masked.

The segments with low S/N that are not used are all of channel 4 for the stars and channel 1 for the asteroids.
The stars have low S/N in channel 4 as this channel has low throughput \citep{Argyriou23} and their spectral energy distributions (SEDs) are declining rapidly.
The asteroids have low S/N in channel 1 as their SEDs are peaked at longer MIR wavelengths.

Additional masking is done for the stars at wavelengths with strong absorption or emission lines.
For A stars, regions near the many hydrogen lines are masked $\pm 0.02~\micron$ from the center of each line.
For G stars, segments 1A and 1B are not used as they include a large number of molecular lines that are challenging to model accurately.
For the other segments, a small number of lines were visually identified and masked $\pm 0.04~\micron$ from their centers.
For O stars, all hydrogen lines $\geq 6.5~\micron$ and a select set of helium lines were masked $\pm 0.02~\micron$ of their centers.
The hydrogen lines in O stars are generally weak below $6.5~\micron$ and fairly well modeled enabling the O stars to fill in the line regions masked for the A stars in segments 1A and 1B.
For segment 1C, the G stars fill in the masked line regions and for segments 2A--3C, the asteroids and G stars fill in the masked line regions.

\begin{figure*}[tbp]
\epsscale{1.2}
\plotone{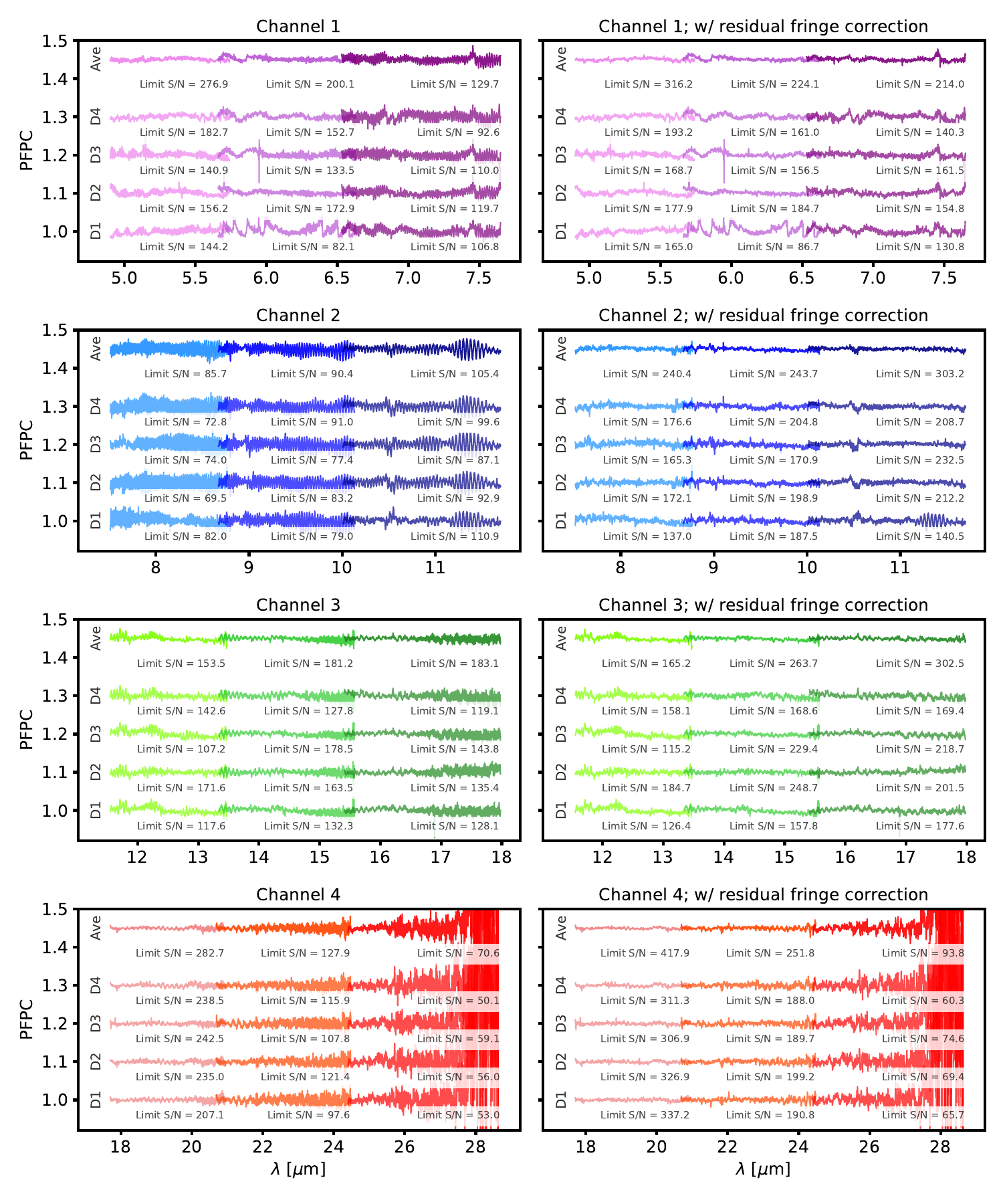}
\caption{The left panels give the derived PFPCs for each dither and the average of the 4 dithers at the top of each plot.
The portion of the PFPCs that is not due to the expected residual fringes is shown in right panels where the PFPCs have been corrected using the pipeline residual fringe correction.
The average S/N over each segment is given below that segment.
\label{fig_pfpc}}
\end{figure*}

The PFPC measurements from each star and asteroid are then combined with sigma clipping to create the average PFPC for each segment and dither.
This is illustrated in Fig.~\ref{fig_pfpc_stability} for one band in each MRS channel.
The PFPCs from each star and asteroid for each dither are overplotted, show significant structure at the expected level of 1-2\%, and show good agreement.
The sigma clipped average for each dither position is shown directly above each dither stack in black.
The models are plotted at the top of each panel and visually illustrate the strong absorption lines that were masked.
The measurements removed by the sigma clipping are plotted fainter than the measurements used in the average.
There were strong spikes in the PFPC averages that were not removed by sigma clipping due to these spikes being present in multiple, but not all of the measurements at a specific wavelength.
These spikes are due to bad pixels that affect only some of the observations and were not masked in the pipeline processing.
These spikes were removed by visually determining the observations affected and masking them.
Combining measurements from the stars and asteroids results in high S/N PFPC measurements at all MRS wavelengths.

This procedure was done for all 12 segments and 4 dither positions giving 48 PFPCs that are plotted in left panels of Fig.~\ref{fig_pfpc}.
The PFPCs show strong variation with channel and dither position.
In channel 1, the dither position variation is quite marked.
Channel 2 and segment 4C show the largest amplitude fixed pattern noise.
The portion of the PFPC that is not due to residual fringes is shown in the right panels of this figure where the pipeline 1d residual fringe correction has been applied to the PFPCs.
While the fixed pattern noise is reduced by the residual fringe correction, there is still significant structure.
This remaining structure is due to sampling artifacts, limited flat field S/N, and interpolation artifacts around strong lines.
The interpolation artifacts are due to the need to interpolate over strong lines when constructing the flat field from observations of planetary nebula NGC 7027 \citep{Law25} with one example due to bright {\sc S IV} emission clearly seen in channel 2 around 10.5~\micron.
Sampling artifacts \citep[see discussion by][]{Law26} are clearly seen in channel 1, dither position 1 where the sharp features are due to pixel-phase effects from part of the PSF falling out of the IFU field of view.
Finally, there are large scale ripples that often vary by dither position at a number of wavelengths (e.g., 5.8, 6.8, 12.3, and 20.2~\micron) whose origins are not clear.

The fixed pattern noise will fundamentally limit the S/N if uncorrected.
This is quantified by measuring the limiting S/N for each segment for each dither and the average of the 4 dither positions both for the PFPCs as measured and the PFPCs after residual fringe correction.
The limiting S/N values are given below each segment in Fig.~\ref{fig_pfpc}.
The limiting S/N on the average provides the best measure of the maximum S/N of MRS spectra.
For sources with emission or molecular lines where the residual fringe correction cannot be applied, the limiting S/N ranges from 100--300 (left panels).
Where the residual fringe correction can be applied, the limiting S/N ranges from 180--430 (right panels).
These are averages over a segment, there will be regions inside of segments that will have lower and higher limiting S/N.

\section{Results}
\label{sec_results}

\subsection{Using PFPCs}

Applying the PFPCs to observations requires non-standard pipeline processing to obtain extracted spectra for individual exposures along with custom post-processing steps.
The detailed method is given below and applies to version 2.0.0 of the jwst pipeline.
First, the observations are reduced using the steps detailed in section~\ref{sec_data}.
The output of the reductions are spectra for each dither position and MRS segment.
These segments are post-processed through the following steps.

\begin{enumerate}
\item Apply the PFPCs for each dither position and segment.
\item Multiplicatively correct the overall level of each dither position to the average of the 4 dithers to remove any small offsets.
\item Average the 4 dither positions after sigma clipping at each wavelength.
The result is given in the `FLUX' column of the output file with the associated `WAVELENGTH' column.
The standard deviation of the mean provides the empirical estimate of the uncertainties and this is given in the `FLUX\_ERROR' column.
\item Run the jwst pipeline residual fringe correction step on the average to remove any remaining residual fringes.
The result in the `RF\_FLUX' column of the output file.
\end{enumerate}
Steps 1--4 are performed by using the MRS-PFPC \texttt{pfpc\_cor} command.  The final 1D spectra have a similar format as the pipeline produced `x1d` files, with the main difference being only having columns for the wavelengths, flux, flux uncertainty, and residual fringe corrected flux.

\begin{figure*}[tbp]
\epsscale{1.0}
\plotone{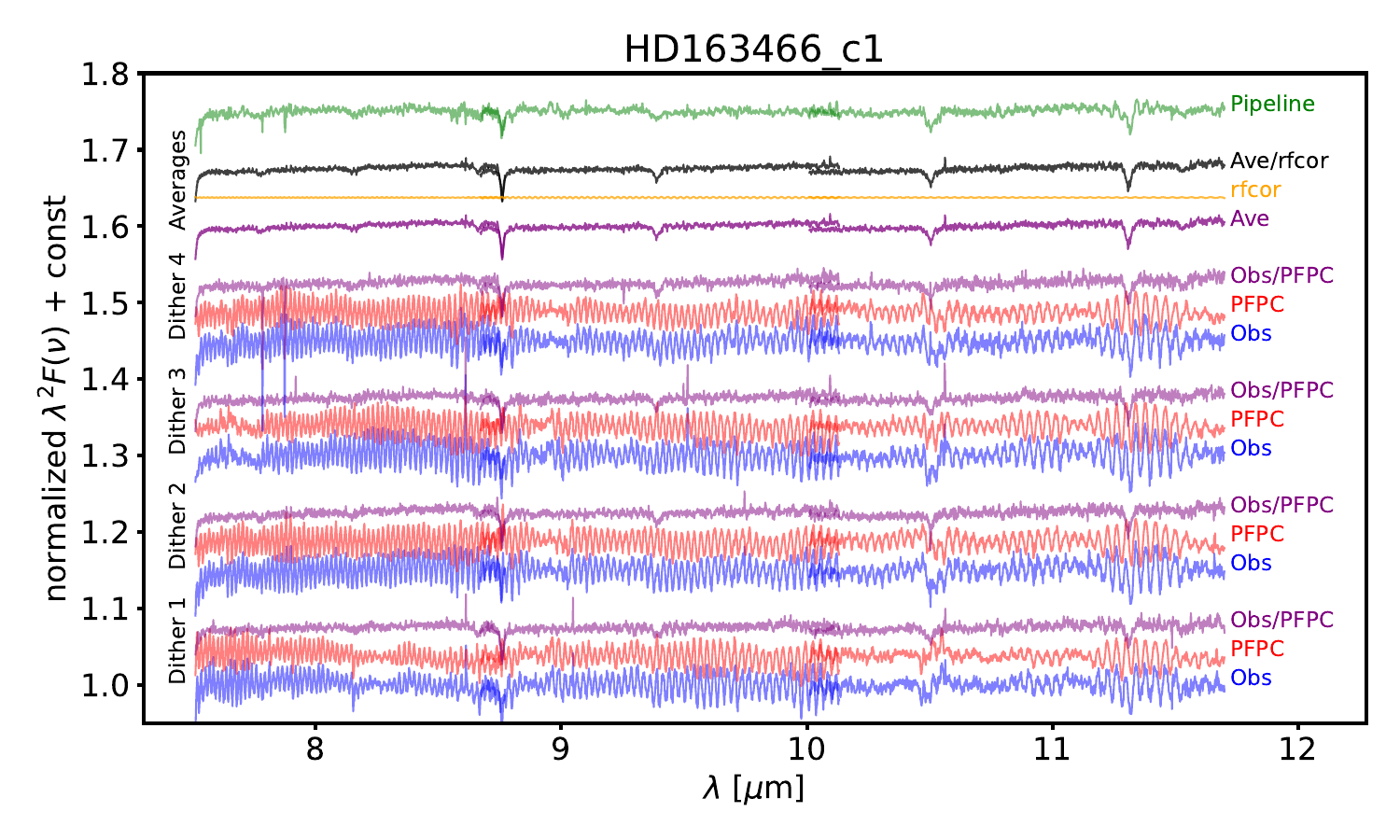}
\caption{The application of the PFPC is illustrated for all 3 bands of channel 2.
HD~163466 is a MRS flux calibration star that was not used in constructing the PFPC.
The extracted spectra for each dither (Obs), PFPC, and division of the two illustrate the stability of the PFPC.
The average of the Obs/PFPC, residual fringe correction for the average (rfcor), and the resulting residual fringe corrected average (Ave/rfcor) are shown.
For reference, the jwst pipeline average spectrum including the residual fringe correction is also plotted (Pipeline).
\label{fig_pfpc_example}}
\end{figure*}

The application of the PFPC to the observation of the A6V flux calibration star HD~163466 (PID: 1536) for all three bands in channel 2 is illustrated in Fig.~\ref{fig_pfpc_example}.
This observation was not used in constructing the PFPCs.
The individual dither spectra for HD~163466 have very similar structures to that seen in the PFPCs.
The spectra divided by the PFPCs show much less structure and are very similar between dither positions.
The sigma clipped average removes most, but not all of the bad pixels seen in the different dither positions and the residual fringe correction is small indicating that the residual fringes were effectively removed with the PFPCs.
The default pipeline spectrum with residual fringe correction is shown and has significant fixed pattern structure that is not seen in the PFPC corrected spectra.
Another marked difference is in profiles of the absorption lines and this can be traced to the fixed pattern noise removed with the PFPCs (e.g., near 10.5~\micron).
The extra fixed pattern noise near the stellar lines (e.g., hydrogen) is likely due to interpolation done over these regions in the creation of the MRS flat field \citep{Law25}.
Finally, the sigma clipped average of the 4 dither positions generally has fewer bad pixels than the default pipeline spectra.
This plot is produced automatically when the MRS-PFPC \texttt{pfpc\_plot} command is used.

\subsection{Spectral artifact removal}

\begin{figure*}[tbp]
\epsscale{1.0}
\plotone{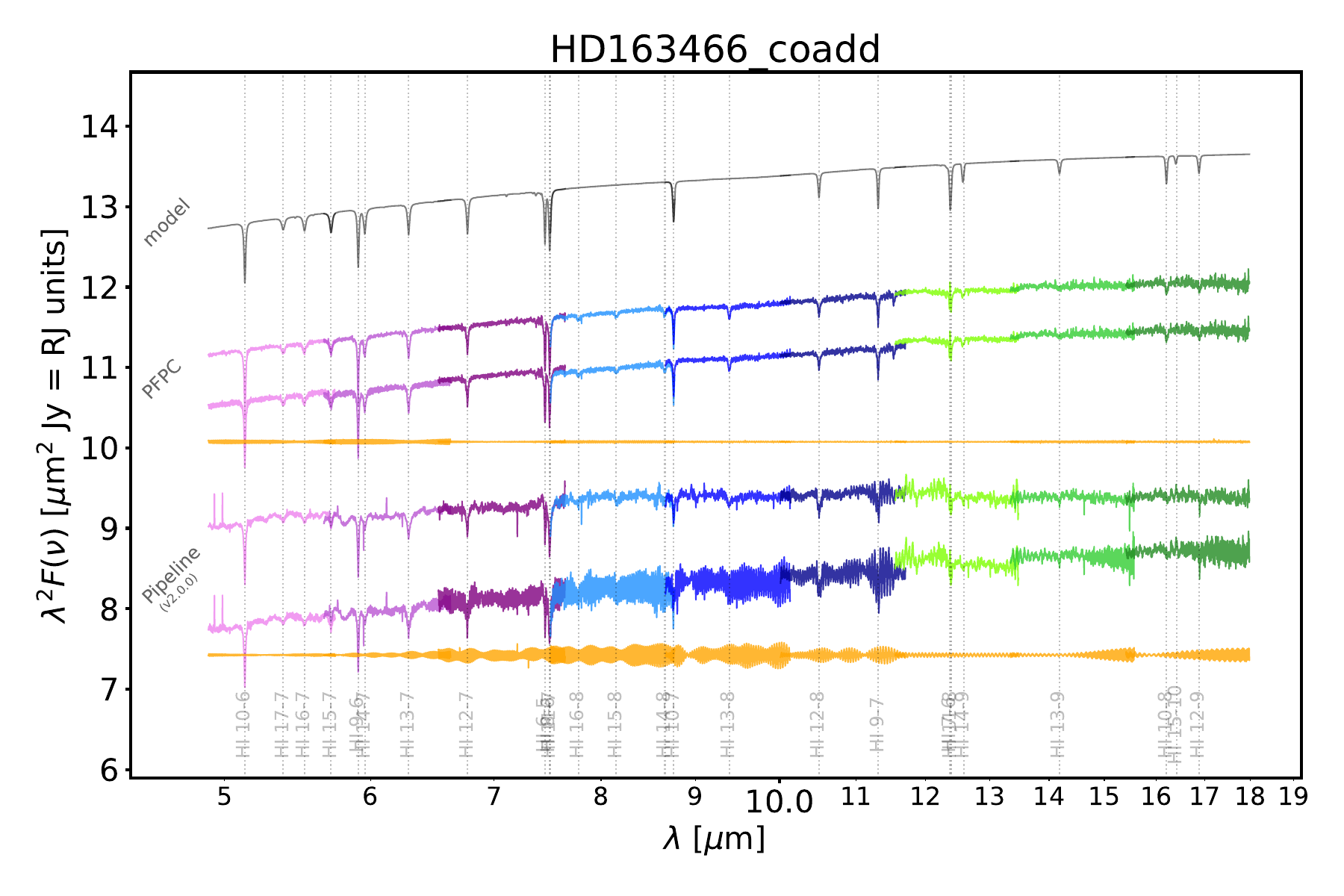}
\caption{The coadded spectra for HD~163466 are plotted as pairs for the pipeline and PFPC.
For each pair of spectra the lower gives the default and the upper the residual fringe and spectral overlap corrected spectra.
The orange line right below each pair gives the residual fringe signal as determined with the residual fringe correction algorithm.
The CALSPEC model for this star is given at the top with the approximate spectral resolution of each segment.
Channel 4 is not shown as it has very low S/N.
\label{fig_spec_shape}}
\end{figure*}

The PFPCs make a significant improvement to the spectra as illustrated in Fig.~\ref{fig_spec_shape}.
Spectra of HD~163466 are shown based on the coaddition of 25 epochs\footnote{The 7th epoch observations in cycle 2 are not included in this total nor used in this paper as they displayed noticeably lower S/N than the other epochs especially in channel 1.  The origin of this difference is under investigation.} of observations taken in cycles 1--4 (PIDs: 1536, 4499, 6607, and 7671).
HD~163466 is a star that is regularly observed to monitor the MRS flux calibration.
Coadding all the epochs results in a high S/N spectrum allowing for spectral artifacts to be more easily seen.
A single epoch of observations show the same behavior as the coadded spectra.
For reference, the CALSPEC model for this star is shown at the top of the plot convolved to the spectral resolution of each segment.
The CALSPEC model was derived by fitting HST/STIS spectroscopy and ground- and space-based photometry using no JWST data \citep{Bohlin14, Gordon22}.
This plot can be made with the MRS-PFPC \texttt{pfpc\_plot} command.

The lower pair of spectra show the pipeline results without residual fringe correction (bottom) and with residual fringe and segment overlap correction (top).
The segment overlap correction is where the regions between segments have been corrected to have the same value.
All the segments are corrected assuming segment 1A is correct.
The need for such a spectral segment correction can be caused by small mismatches in the flux calibration between segments that may be caused by uncertainties in the time-dependent throughput corrections.
The residual fringe correction applied is shown as the orange line below the pair.
For the pipeline spectra, the segment overlap correction ranges from 0.91 to 1.03.
The overall shape of the segment overlap corrected pipeline is flatter than the default pipeline spectra and than the CALSPEC model.
In addition to the overall shape, the corrected pipeline spectrum has a number of spectral artifacts seen mainly as ``wiggles" in the continuum.

The upper pair of spectra show the PFPC results without residual fringe correction (bottom) and with residual fringe and segment overlap correction (top).
In contrast to the pipeline spectra, both PFPC spectra have the similar spectral shapes with segment overlap corrections 0.96 to 1.0 and have a similar spectral shape as the CALSPEC model.
The residual fringe correction is shown below the pair and is much smaller than the same correction needed for the pipeline spectrum.
The PFPC spectra visually show significantly higher S/N, have many fewer spectral artifacts, and are more similar to the CALSPEC model than the pipeline spectra.
The S/N performance is quantified in Sec.~\ref{sec_sn}.
The line depths in the model spectrum and the PFPC spectra are approximately the same, with the model having stronger lines at the longer wavelengths.
The HI 7-6 line at 13.37~\micron\ shows emission in its core that is not seen in the model.
This emission is seen in the pipeline PFPC spectra before residual fringe correction and so is not due to any of the corrections applied.
Such emission at varying levels is seen for the other A dwarf stars, but not for the G dwarf stars.
This may be additional evidence for non-LTE effects in A dwarfs \citep{Hubeny81, Mashonkina20} that are generally modeled assuming LTE \citep{Bohlin17}.

\begin{figure}[tbp]
\epsscale{1.15}
\plotone{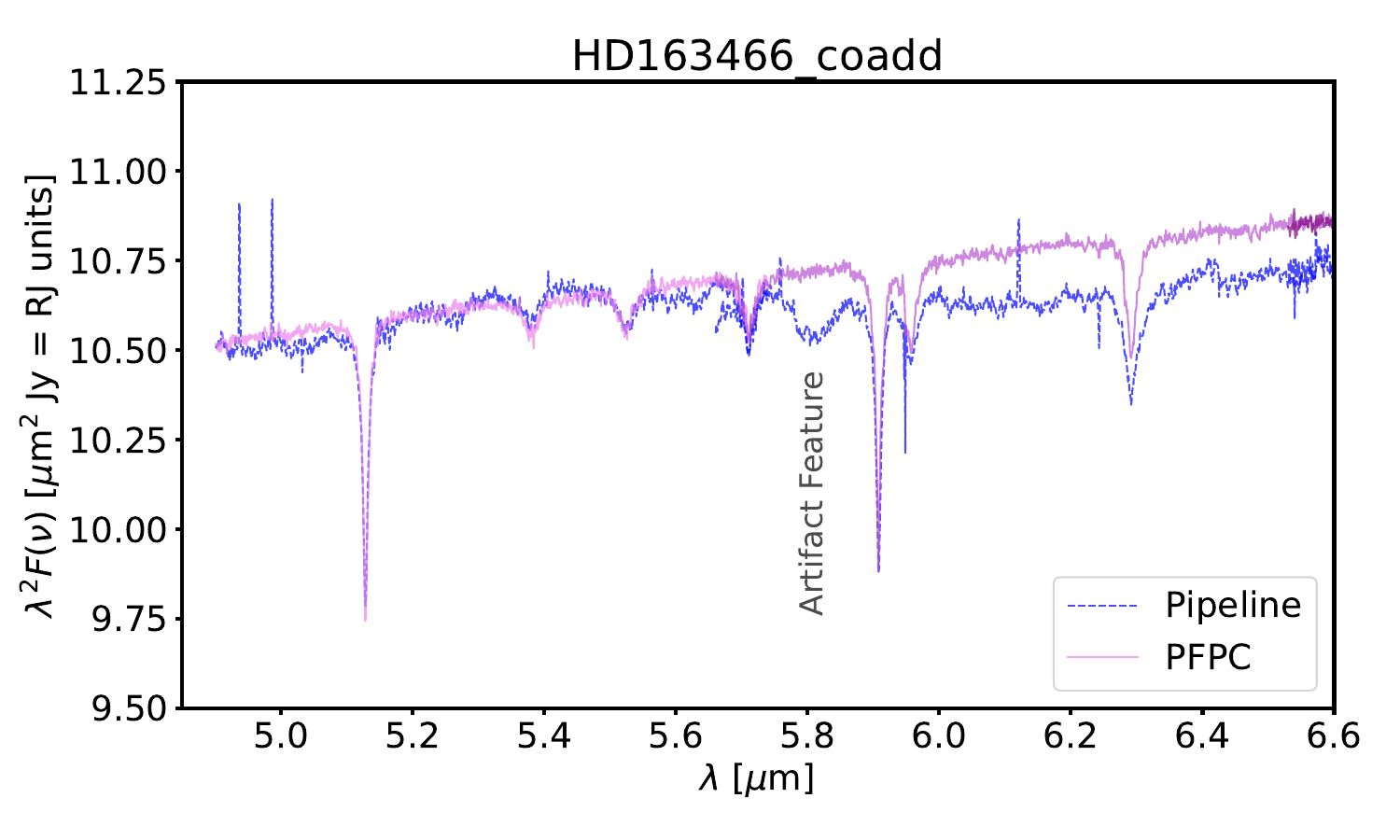}
\caption{Segments 1A and 1B are shown for the pipeline and PFPC spectra.
Both have been had residual fringe and spectral overlap correction applied.
The spectral artifact at 5.8~\micron\ is clearly present in the pipeline spectrum and absent in the PFPC spectrum.
\label{fig_58_artifact}}
\end{figure}

One particular spectral artifact that is removed by the PFPC correction is one seen at 5.8~\micron.
The pipeline and PFPC corrected spectra are plotted for the region around 5.8~\micron\ in Fig.~\ref{fig_58_artifact}.
The pipeline spectra clearly show a broad, symmetric feature at 5.8~\micron\ that has the properties of a carbonyl feature due to C=O bonds in dust grains \citep{Pendleton25}.
This sightline has very little dust, hence it is surprising to see this feature.
The PFPC corrected spectrum does not show this feature at all, indicating that this is a fully instrumental spectral artifact.
The instrumental origin of this feature is also supported by the variations in the PFPC corrections for this segment (see Fig.~\ref{fig_pfpc_stability}) where dither 1 shows artifacts that likely contribute to this feature, dither 2 does not show this feature, and dithers 3 and 4 clearly show this feature.
This detailed comparison shown in Fig.~\ref{fig_58_artifact} also illustrates that the PFPC removes multiple smaller spectral artifacts resulting in a spectrum that more closely matches that expected for an A dwarf star.

\subsection{S/N improvement}
\label{sec_sn}

\begin{figure*}[tbp]
\epsscale{1.15}
\plotone{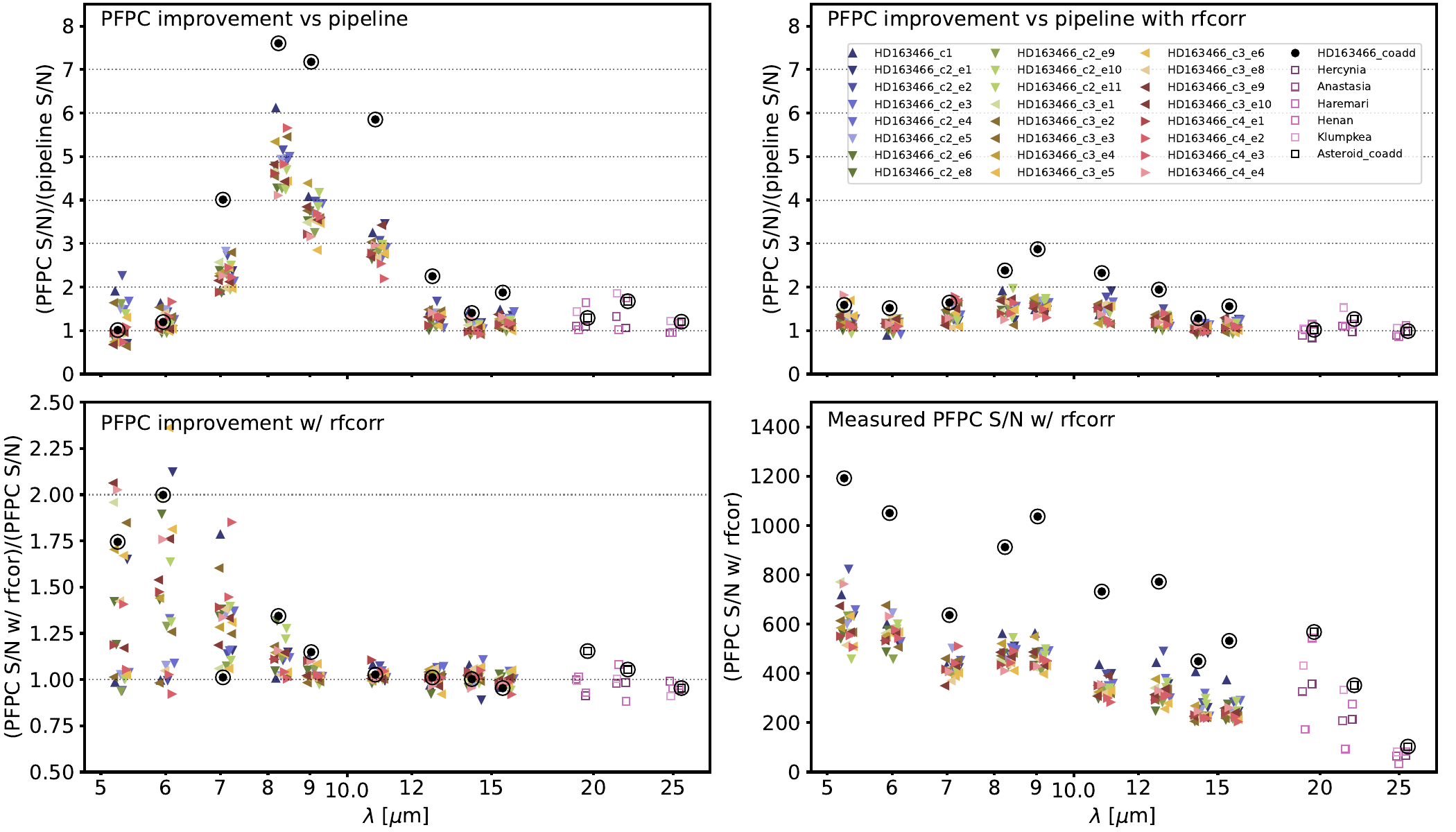}
\caption{The improvement in channels 1--3 in the S/N with the application the PFPCs is shown for the 25 epochs of HD~163466 observations and their coadd (circled points).
For channel 4 where this star has very low S/N, the improvement is shown using five asteroids and their coadd after dividing by a quadratic fit for each segment.
Neither HD~163466 nor the asteroids were used in the creation of the PFPCs.
The S/N is measured using visually clean spectral regions for each spectral segment.
The wavelengths of the points have been shifted randomly by small amounts for display purposes.
The top left panel shows the PFPC improvement with no residual fringe correction done.
For sources where the residual fringe correction can be done, the top right panel shows the improvement with the PFPCs.
The bottom left panel gives the improvement for the PFPC corrected spectra when using the residual fringe correction.
The achieved S/N values for the PFPC and residual fringe corrected spectra are plotted in the lower right panel.
\label{fig_pfpc_performance}}
\end{figure*}

To measure the performance of the PFPCs, spectral regions in each segment were identified that have no expected spectral lines from stellar models and visually appeared to be clean.
The S/N in each of these regions was measured by fitting a line to the continuum and measuring the standard deviation.
Fig.~\ref{fig_pfpc_performance} shows the S/N measurements plotted versus wavelength for the cycles 1--4 observations of HD~163466 and the coaddition of the 25 epochs for channels 1--3.
Five asteroids (PID: 2361) were used for channel 4 as they have much higher S/N than stars.
The asteroid coaddition was done by first dividing a quadratic fit to each segment for each asteroid and then coadding the resulting normalized spectra.
The repeated observations of the star and the set of asteroids provide a measure of how variations in the exact pointing and grating wheel repeatability impact the fixed pattern noise.

The top left panel gives the S/N improvement compared to the pipeline spectra from applying the PFPCs as a function of wavelength.
These improvements are applicable for sources where residual fringe correction cannot be applied as the source has too many regularly-spaced lines (e.g., molecular or emission).
The improvement is largest in channel 2 and segment 1C with factors of 2--6 improvements in the S/N for single observations with larger improvements seen for the coadded spectra.
This is as expected as the fixed pattern noise is largest for channel 1C and all of channel 2 (see Fig.~\ref{fig_pfpc}).
For other segments, the S/N improvement factor is between 1 and 2.

For sources where the residual fringe correction can be applied, the top right panel shows that the PFPC S/N improvements compared the the pipeline are generally a factor between 1 and 2 with channel 2 showing the largest improvements.
The coadded spectra show larger improvements indicating that the improvement is larger when the photon noise S/N is higher.
The impact of the residual fringe correction on the PFPC corrected spectra is shown in the bottom left panel.
In this case, the improvement is up to a factor of two in channel 1 decreasing rapidly through channel 2 and showing no significant improvement at longer wavelengths, expect for segment 4A.
For some epochs the improvement in channels 1 neglibable.
This likely indicates that the residual fringes that are being corrected the fringes varying due to sub-pixel differences between the actual and average pixel dither locations and the small non-repeatability of the grating wheels.
The shorter wavelengths appear to be be the most sensitive to such positional differences.

The delivered S/N values for the PFPC and residual fringe corrected spectra are given in the bottom right panel.
The S/N varies significantly between different epochs indicating there are sources of noise that are not captured in the corrections done.
The coadded spectra show higher S/N as expected, but not as high if the only noise was photon noise.
For example, channel 1A has a single epoch S/N around 600 and the coadded S/N is 1200.
This is a factor of two better, yet if the noise was only due to photon noise then the coadded spectra should be a factor of five higher.
This indicates that the coadded spectra S/N are likely limited by the S/N of the PFPCs themselves and the coadded S/N gives the PFPC S/N.
This emphasizes that the maximum S/N possible is set by whichever is lowest, the photon S/N of the observations or the S/N of the PFPC.
As a test that the PFPC correction S/N is the limiting factor, the cycles 1 and 2 HD~163466 observations were added to the set used to create the PFPCs and these new PFPCs were used to correct the 14 HD~163466 observations from cycles 3 and 4.
The coadd of these 14 epochs showed a higher S/N than the original coadd of 25 epochs indicating that with more calibration observations, the PFPC S/N does increase.
For the final PFPCs, the HD~163466 observations were not used as the PFPCs would be strongly biased to the observations of this one star.
Having a diversity of sources for the PFPC measurements is important to average over model uncertainties and fill in regions of strong lines or low S/N.

The HD~163466 points from different cycles are shown with different symbols to allow for trends with time to be investigated.
The time-dependent throughput loss \citep{Law25} is corrected by the default pipeline, but potentially could affect the PFPC causing it to be time-dependent as well.
However, the results from different cycles do not show any obvious trends as the symbols are well mixed both in the measured S/N and the correction factors.
This indicates both that the achieved S/N has no significant time-dependence in Channels 1-3, and that the variations between observations are caused by sources not corrected for in the PFPCs (e.g., variable cosmic ray rate).

\section{Summary}
\label{sec_summary}

Corrections for the fixed pattern noise in MIRI MRS spectra were derived for all 4 channels, 3 grating settings, and 4 dither positions for the default dither pattern (4-POINT, NEGATIVE, POINT-SOURCE, and ALL\_MRS).
Observations of 2 O dwarfs, 3 A dwarfs, and 4 G dwarfs were used along with 2 asteroids providing high S/N coverage over the full MRS wavelength range.
In particular, the combination of stars and asteroids allowed for good corrections around the stellar absorption and emission lines.

The PFPCs display significant wavelength structure that varies between dither positions, channels, and gratings.
The fixed pattern noise is due to a number of sources including residual fringes, sampling artifacts, and flat field noise.
The wavelength dependence of the fixed pattern noise is highly variable displaying both narrow and broad structures.
Channel 1 shows the largest PFPC variations between dither positions, channel 2 shows the largest amplitude PFPCs, and channels 3 and 4 show the smallest PFPC corrections overall except for the longest wavelength of channel 4 that may be impacted by measurement S/N in the observations used.

Without the use of the PFPCs, the limiting S/N of MRS spectra was quantified by measuring the S/N of the PFPCs with values ranging from 100--300 depending on the specific spectral segment. 
These S/N values are applicable to targets where the pipeline residual fringe correction cannot be used due to the targets having structures that mimic the residual fringes (e.g., molecular lines or many emission lines).
For targets where the pipeline residual fringe correction can be applied, the limiting S/N is 180--430 and this was measured using the PFPCs after correcting them with the pipeline residual fringe correction.

With the use of the PFPCs, the MRS limiting S/N is significantly higher with values exceeding 1000 at the shorter wavelengths for targets where the residual fringe correction can also be applied.
This was measured using many epochs of observations of one A dwarf that is observed multiple times during each year to monitor the MRS throughput.
Leveraging the many observations, it is most likely that the S/N limit is due to the S/N of the PFPC and not the observations themselves.
The improvement using the PFPCs is largest for the case where the residual fringe correction cannot be applied, seeing increases in the S/N of up to almost 8$\times$.
When the residual fringe correction can be used, the improvements are still quite significant showing values up to 3$\times$.

Not only is the S/N improved, but there are a number of broad spectral artifacts that are removed by the PFPCs.
The spectral artifacts removed include both clear isolated features and overall structure that affects the continuum shape.
An example of a specific feature that is removed is a broad symmetric feature at 5.8~\micron\ that mimics that expected from dust grains.
The overall spectral structures are such that when the spectral overlaps between segments are used to adjust each segment for agreement, the spectral shape is distorted.
After the PFPCs are applied, spectral overlap correction is smaller and the spectral shape is preserved.
Correcting for these broad spectral artifacts is important for observations at almost any S/N.

The PFPCs are available for use via the MRS-PFPC python repository\footnote{\url{https://github.com/STScI-MIRI/MRS-PFPC}}.
Before applying the PFPCs, the jwst pipeline is run on the raw/uncal data via the supplied \texttt{pfpc\_proc} command to produce the specific pipeline outputs required for each dither and segment.
These outputs are corrected using the PFPCs with the \texttt{pfpc\_cor} command and dither combined spectra for each segment are created without and with residual fringe correction.
The pipeline and PFPC spectra can be visualized with the \texttt{pfpc\_plot} command.

The PFPCs are directly coupled to the version of the jwst pipeline used as the fixed pattern noise is directly tied to the specific reference files used (e.g., flat fields) and algorithms in the pipeline steps used.
Thus, as the jwst pipeline algorithms and reference files are updated new PFPCs will be necessary.
The MRS-PFPC package will be updated to include new version of the PFPCs and the MRS-PFPC \texttt{pfpc\_cor} code will automatically pick the appropriate PFPC version.

Future work could included expanding the PFPCs to other dither patterns and enhancing the PFPCs to account for small offsets from the nominal dither positions.
The expansion to other dither patterns is dependent on having appropriate observations of stars and asteroids.
Accounting for offsets from the nominal dither patterns could be done by including the existing and future observations taken with mini-grids around each dither position \citep{Gasman23, Gasman24}.
Such an expansion would be particularly useful for channels 1 and 2 where there is clear evidence for significant residual fringe variation likely with small pointing offsets.
If and how the PFPCs could be incorporated into the JWST pipeline is another area for future study.

\begin{acknowledgements}
We thank the MIRI instrument team for fruitful discussions, especially Greg Sloan.
\end{acknowledgements}

\facilities{JWST (MIRI)}

This work is based on observations made with the NASA/ESA/CSA James Webb Space Telescope.
The data were obtained from the Mikulski Archive for Space Telescopes at the Space Telescope Science Institute, which is operated by the Association of Universities for Research in Astronomy, Inc., under NASA contract NAS 5-03127 for JWST. These observations are associated with program \# 1536, 1538, 1549, 2361, 3779, 4496, 4497, and 6604.
The data used in this paper can be found in MAST: \dataset[10.17909/jtxf-pb27]{http://dx.doi.org/10.17909/jtxf-pb27}.
The code used for this paper is available\footnote{\url{https://github.com/STScI-MIRI/MRS-PFPC}} \citep{mrspfpc}.

\software{\texttt{astropy} \citep{astropy:2013,astropy:2018,astropy:2022}, \texttt{matplotlib} \citep{Hunter:2007}, \texttt{numpy} \citep{numpy}, \texttt{photutils} \citep{Photutils_17129028}, \texttt{python} \citep{python}, \texttt{scipy} \citep{2020SciPy-NMeth,scipy_18736568}, \texttt{jwst} \citep{jwst_17515973}, \texttt{scikit-image} \citep{scikit-image}, and \texttt{\href{https://www.tomwagg.com/software-citation-station/}{The Software Citation Station}} \citep{software-citation-station-paper,software-citation-station-zenodo}.}

\typeout{}
\bibliography{dust}{}

@ARTICLE{Licardro12,
       author = {{Licandro}, J. and {Hargrove}, K. and {Kelley}, M. and {Campins}, H. and {Ziffer}, J. and {Al{\'\i}-Lagoa}, V. and {Fern{\'a}ndez}, Y. and {Rivkin}, A.},
        title = "{5-14 {\ensuremath{\mu}}m Spitzer spectra of Themis family asteroids}",
      journal = {\aap},
         year = 2012,
        month = jan,
       volume = {537},
          eid = {A73},
        pages = {A73},
          doi = {10.1051/0004-6361/201118142},
       adsurl = {https://ui.adsabs.harvard.edu/abs/2012A&A...537A..73L}
}

@ARTICLE{Hubeny81,
       author = {{Hubeny}, I.},
        title = "{Non-LTE analysis of the ultraviolet spectrum of A type stars. II Theoretical considerations and interpretation of the VEGA Lyman-alpha region}",
      journal = {\aap},
         year = 1981,
        month = may,
       volume = {98},
       number = {1},
        pages = {96-111},
       adsurl = {https://ui.adsabs.harvard.edu/abs/1981A&A....98...96H}
}

@ARTICLE{Bohlin17,
       author = {{Bohlin}, Ralph C. and {M{\'e}sz{\'a}ros}, Szabolcs and {Fleming}, Scott W. and {Gordon}, Karl D. and {Koekemoer}, Anton M. and {Kov{\'a}cs}, J{\'o}zsef},
        title = "{A New Stellar Atmosphere Grid and Comparisons with HST/STIS CALSPEC Flux Distributions}",
      journal = {\aj},
         year = 2017,
        month = may,
       volume = {153},
       number = {5},
          eid = {234},
        pages = {234},
          doi = {10.3847/1538-3881/aa6ba9},
archivePrefix = {arXiv},
       eprint = {1704.00653},
 primaryClass = {astro-ph.SR},
       adsurl = {https://ui.adsabs.harvard.edu/abs/2017AJ....153..234B}
}

@ARTICLE{Mashonkina20,
       author = {{Mashonkina}, L. and {Ryabchikova}, T. and {Alexeeva}, S. and {Sitnova}, T. and {Zatsarinny}, O.},
        title = "{Chemical diversity among A-B stars with low rotational velocities: non-LTE abundance analysis}",
      journal = {\mnras},
         year = 2020,
        month = dec,
       volume = {499},
       number = {3},
        pages = {3706-3719},
          doi = {10.1093/mnras/staa3099},
archivePrefix = {arXiv},
       eprint = {2010.02058},
 primaryClass = {astro-ph.SR},
       adsurl = {https://ui.adsabs.harvard.edu/abs/2020MNRAS.499.3706M}
}

@ARTICLE{Rigby23,
       author = {{Rigby}, Jane and {Perrin}, Marshall and {McElwain}, Michael and {Kimble}, Randy and {Friedman}, Scott and {Lallo}, Matt and {Doyon}, Ren{\'e} and {Feinberg}, Lee and {Ferruit}, Pierre and {Glasse}, Alistair and et al.},
        title = "{The Science Performance of JWST as Characterized in Commissioning}",
      journal = {\pasp},
         year = 2023,
        month = apr,
       volume = {135},
       number = {1046},
          eid = {048001},
        pages = {048001},
          doi = {10.1088/1538-3873/acb293},
archivePrefix = {arXiv},
       eprint = {2207.05632},
 primaryClass = {astro-ph.IM},
       adsurl = {https://ui.adsabs.harvard.edu/abs/2023PASP..135d8001R}
}

@ARTICLE{Heap95,
       author = {{Heap}, S.~R. and {Brandt}, J.~C. and {Randall}, C.~E. and {Carpenter}, K.~G. and {Leckrone}, D.~S. and {Maran}, S.~P. and {Smith}, A.~M. and {Beaver}, E.~A. and {Boggess}, A. and {Ebbets}, D.~C. and et al.},
        title = "{The Goddard High Resolution Spectrograph: In-Orbit Performance}",
      journal = {\pasp},
         year = 1995,
        month = sep,
       volume = {107},
        pages = {871},
          doi = {10.1086/133635},
       adsurl = {https://ui.adsabs.harvard.edu/abs/1995PASP..107..871H}
}

@ARTICLE{Leckrone89,
       author = {{Leckrone}, David S. and {Adelman}, Saul J.},
        title = "{On the Signal-to-Noise Ratio in IUE High-Dispersion Spectra}",
      journal = {\apjs},
         year = 1989,
        month = oct,
       volume = {71},
        pages = {387},
          doi = {10.1086/191379},
       adsurl = {https://ui.adsabs.harvard.edu/abs/1989ApJS...71..387L}
}

@software{Bushouse26,
author = {Bushouse, Howard and Eisenhamer, Jonathan and Dencheva, Nadia and Davies, James and Greenfield, Perry and Morrison, Jane and Hodge, Phil and Simon, Bernie and Grumm, David and Droettboom, Michael and Slavich, Edward and Sosey, Megan and Pauly, Tyler and Miller, Todd and Jedrzejewski, Robert and Hack, Warren and Davis, David and Crawford, Steven and Law, David and Gordon, Karl and Regan, Michael and Cara, Mihai and MacDonald, Ken and Bradley, Larry and Shanahan, Clare and Jamieson, William and Teodoro, Mairan and Williams, Thomas and Pena-Guerrero, Maria and Graham, Brett and Molter, Edward and Brandt, Timothy and Hayes, Christian and Cooper, Rachel and Clarke, Melanie and Filippazzo, Joseph},
doi = {10.5281/zenodo.7038885},
month = apr,
title = {{JWST Calibration Pipeline}},
url = {https://github.com/spacetelescope/jwst},
version = {2.0.0},
year = {2026}
}

@ARTICLE{Rieke15,
       author = {{Rieke}, G.~H. and {Wright}, G.~S. and {B{\"o}ker}, T. and {Bouwman}, J. and {Colina}, L. and {Glasse}, Alistair and {Gordon}, K.~D. and {Greene}, T.~P. and {G{\"u}del}, Manuel and {Henning}, Th. and {Justtanont}, K. and {Lagage}, P. -O. and {Meixner}, M.~E. and {N{\o}rgaard-Nielsen}, H. -U. and {Ray}, T.~P. and {Ressler}, M.~E. and {van Dishoeck}, E.~F. and {Waelkens}, C.},
        title = "{The Mid-Infrared Instrument for the James Webb Space Telescope, I: Introduction}",
      journal = {\pasp},
         year = 2015,
        month = jul,
       volume = {127},
       number = {953},
        pages = {584},
          doi = {10.1086/682252},
archivePrefix = {arXiv},
       eprint = {1508.02294},
 primaryClass = {astro-ph.IM},
       adsurl = {https://ui.adsabs.harvard.edu/abs/2015PASP..127..584R}
}

@ARTICLE{Wright23,
       author = {{Wright}, Gillian S. and {Rieke}, George H. and {Glasse}, Alistair and {Ressler}, Michael and {Garc{\'\i}a Mar{\'\i}n}, Macarena and {Aguilar}, Jonathan and {Alberts}, Stacey and {{\'A}lvarez-M{\'a}rquez}, Javier and {Argyriou}, Ioannis and {Banks}, Kimberly and et al.},
        title = "{The Mid-infrared Instrument for JWST and Its In-flight Performance}",
      journal = {\pasp},
         year = 2023,
        month = apr,
       volume = {135},
       number = {1046},
          eid = {048003},
        pages = {048003},
          doi = {10.1088/1538-3873/acbe66},
       adsurl = {https://ui.adsabs.harvard.edu/abs/2023PASP..135d8003W}
}

@ARTICLE{Wells15,
       author = {{Wells}, Martyn and {Pel}, J.-W. and {Glasse}, Alistair and {Wright}, G.~S. and {Aitink-Kroes}, Gabby and {Azzollini}, Ruym{\'a}n and {Beard}, Steven and {Brandl}, B.~R. and {Gallie}, Angus and {Geers}, V.~C. and et al.},
        title = "{The Mid-Infrared Instrument for the James Webb Space Telescope, VI: The Medium Resolution Spectrometer}",
      journal = {\pasp},
         year = 2015,
        month = jul,
       volume = {127},
       number = {953},
        pages = {646},
          doi = {10.1086/682281},
archivePrefix = {arXiv},
       eprint = {1508.03070},
 primaryClass = {astro-ph.IM},
       adsurl = {https://ui.adsabs.harvard.edu/abs/2015PASP..127..646W}
}

@ARTICLE{Argyriou23,
       author = {{Argyriou}, Ioannis and {Glasse}, Alistair and {Law}, David R. and {Labiano}, Alvaro and {{\'A}lvarez-M{\'a}rquez}, Javier and {Patapis}, Polychronis and {Kavanagh}, Patrick J. and {Gasman}, Danny and {Mueller}, Michael and {Larson}, Kirsten and et al.},
        title = "{JWST MIRI flight performance: The Medium-Resolution Spectrometer}",
      journal = {\aap},
         year = 2023,
        month = jul,
       volume = {675},
          eid = {A111},
        pages = {A111},
          doi = {10.1051/0004-6361/202346489},
archivePrefix = {arXiv},
       eprint = {2303.13469},
 primaryClass = {astro-ph.IM},
       adsurl = {https://ui.adsabs.harvard.edu/abs/2023A&A...675A.111A}
}

@ARTICLE{Bohlin14,
       author = {{Bohlin}, Ralph C. and {Gordon}, Karl D. and {Tremblay}, P.-E.},
        title = "{Techniques and Review of Absolute Flux Calibration from the Ultraviolet to the Mid-Infrared}",
      journal = {\pasp},
         year = 2014,
        month = aug,
       volume = {126},
       number = {942},
        pages = {711},
          doi = {10.1086/677655},
archivePrefix = {arXiv},
       eprint = {1406.1707},
 primaryClass = {astro-ph.IM},
       adsurl = {https://ui.adsabs.harvard.edu/abs/2014PASP..126..711B}
}

@ARTICLE{Decleir25,
       author = {{Decleir}, Marjorie and {Gordon}, Karl D. and {Misselt}, Karl A. and {G{\"u}nay}, Burcu and {Roman-Duval}, Julia and {Zeegers}, Sascha T.},
        title = "{A First Taste of MEAD (Measuring Extinction and Abundances of Dust). I. Diffuse Milky Way Interstellar Dust Extinction Features in JWST Infrared Spectra}",
      journal = {\aj},
         year = 2025,
        month = feb,
       volume = {169},
       number = {2},
          eid = {99},
        pages = {99},
          doi = {10.3847/1538-3881/ada147},
archivePrefix = {arXiv},
       eprint = {2412.14378},
 primaryClass = {astro-ph.GA},
       adsurl = {https://ui.adsabs.harvard.edu/abs/2025AJ....169...99D}
}

@ARTICLE{Zeegers25,
       author = {{Zeegers}, S.~T. and {Marshall}, Jonathan P. and {Gordon}, Karl D. and {Misselt}, Karl A. and {Otten}, G.~P.~P.~L. and {Bouwman}, Jeroen and {Chiar}, Jean and {Decleir}, Marjorie and {Dharmawardena}, Thavisha and {Kemper}, F. and et al.},
        title = "{Investigating Silicate, Carbon, and Water in the Diffuse Interstellar Medium: The First Shots from WISCI}",
      journal = {\apj},
         year = 2025,
        month = jul,
       volume = {987},
       number = {1},
          eid = {25},
        pages = {25},
          doi = {10.3847/1538-4357/add73b},
archivePrefix = {arXiv},
       eprint = {2506.20033},
 primaryClass = {astro-ph.GA},
       adsurl = {https://ui.adsabs.harvard.edu/abs/2025ApJ...987...25Z}
}

@ARTICLE{Argyriou20,
       author = {{Argyriou}, Ioannis and {Wells}, Martyn and {Glasse}, Alistair and {Lee}, David and {Royer}, Pierre and {Vandenbussche}, Bart and {Malumuth}, Eliot and {Glauser}, Adrian and {Kavanagh}, Patrick J. and {Labiano}, Alvaro and et al.},
        title = "{The nature of point source fringes in mid-infrared spectra acquired with the James Webb Space Telescope}",
      journal = {\aap},
         year = 2020,
        month = sep,
       volume = {641},
          eid = {A150},
        pages = {A150},
          doi = {10.1051/0004-6361/202037535},
archivePrefix = {arXiv},
       eprint = {2007.16143},
 primaryClass = {astro-ph.IM},
       adsurl = {https://ui.adsabs.harvard.edu/abs/2020A&A...641A.150A}
}

@ARTICLE{Gasman23,
       author = {{Gasman}, Danny and {Argyriou}, Ioannis and {Sloan}, G.~C. and {Aringer}, Bernhard and {{\'A}lvarez-M{\'a}rquez}, Javier and {Fox}, Ori and {Glasse}, Alistair and {Glauser}, Adrian and {Jones}, Olivia C. and {Justtanont}, Kay and {Kavanagh}, Patrick J. and {Klaassen}, Pamela and {Labiano}, Alvaro and {Larson}, Kirsten and {Law}, David R. and {Mueller}, Michael and {Nayak}, Omnarayani and {Noriega-Crespo}, Alberto and {Patapis}, Polychronis and {Royer}, Pierre and {Vandenbussche}, Bart},
        title = "{JWST MIRI/MRS in-flight absolute flux calibration and tailored fringe correction for unresolved sources}",
      journal = {\aap},
         year = 2023,
        month = may,
       volume = {673},
          eid = {A102},
        pages = {A102},
          doi = {10.1051/0004-6361/202245633},
archivePrefix = {arXiv},
       eprint = {2212.03596},
 primaryClass = {astro-ph.IM},
       adsurl = {https://ui.adsabs.harvard.edu/abs/2023A&A...673A.102G}
}

@ARTICLE{Gasman24,
       author = {{Gasman}, Danny and {Argyriou}, Ioannis and {Morrison}, Jane E. and {Law}, David R. and {Glasse}, Alistair and {Gordon}, Karl D. and {Kavanagh}, Patrick J. and {Lage}, Craig and {Patapis}, Polychronis and {Sloan}, Gregory C.},
        title = "{The MIRI/MRS Library. I. Empirically correcting detector charge migration in unresolved sources}",
      journal = {\aap},
         year = 2024,
        month = aug,
       volume = {688},
          eid = {A226},
        pages = {A226},
          doi = {10.1051/0004-6361/202450241},
archivePrefix = {arXiv},
       eprint = {2406.10835},
 primaryClass = {astro-ph.IM},
       adsurl = {https://ui.adsabs.harvard.edu/abs/2024A&A...688A.226G}
}

@ARTICLE{Gasman25,
       author = {{Gasman}, Danny and {Argyriou}, Ioannis and {Law}, David R. and {Glasse}, Alistair and {Gordon}, Karl D. and {Kavanagh}, Patrick J. and {Morrison}, Jane E. and {Patapis}, Polychronis and {Sloan}, Gregory C.},
        title = "{The MIRI/MRS Library: II. Pointing-based defringing of unresolved sources}",
      journal = {\aap},
         year = 2025,
        month = may,
       volume = {697},
          eid = {A58},
        pages = {A58},
          doi = {10.1051/0004-6361/202554055},
       adsurl = {https://ui.adsabs.harvard.edu/abs/2025A&A...697A..58G}
}

@ARTICLE{Law25,
       author = {{Law}, David R. and {Argyriou}, Ioannis and {Gordon}, Karl D. and {Sloan}, G.~C. and {Gasman}, Danny and {Glasse}, Alistair and {Larson}, Kirsten and {Fletcher}, Leigh N. and {Labiano}, Alvaro and {Noriega-Crespo}, Alberto},
        title = "{The James Webb Space Telescope Absolute Flux Calibration. III. Mid-infrared Instrument Medium Resolution Integral Field Unit Spectrometer}",
      journal = {\aj},
         year = 2025,
        month = feb,
       volume = {169},
       number = {2},
          eid = {67},
        pages = {67},
          doi = {10.3847/1538-3881/ad9685},
archivePrefix = {arXiv},
       eprint = {2409.15435},
 primaryClass = {astro-ph.IM},
       adsurl = {https://ui.adsabs.harvard.edu/abs/2025AJ....169...67L}
}

@ARTICLE{Patapis24,
       author = {{Patapis}, Polychronis and {Argyriou}, Ioannis and {Law}, David R. and {Glauser}, Adrian M. and {Glasse}, Alistair and {Labiano}, Alvaro and {{\'A}lvarez-M{\'a}rquez}, Javier and {Kavanagh}, Patrick J. and {Gasman}, Danny and {Mueller}, Michael and {Larson}, Kirsten and {Vandenbussche}, Bart and {Lee}, David and {Klaassen}, Pamela and {Guillard}, Pierre and {Wright}, Gillian S.},
        title = "{Geometric distortion and astrometric calibration of the JWST MIRI Medium Resolution Spectrometer}",
      journal = {\aap},
         year = 2024,
        month = feb,
       volume = {682},
          eid = {A53},
        pages = {A53},
          doi = {10.1051/0004-6361/202347339},
archivePrefix = {arXiv},
       eprint = {2307.01025},
 primaryClass = {astro-ph.IM},
       adsurl = {https://ui.adsabs.harvard.edu/abs/2024A&A...682A..53P}
}

@ARTICLE{Law26,
       author = {{Law}, David R. and {Clarke}, Melanie},
        title = "{Mitigating Resampling Artifacts for the JWST Integral-field Unit Spectrometers with Adaptive Trace Modeling}",
      journal = {\aj},
         year = 2026,
        month = may,
       volume = {171},
       number = {5},
          eid = {304},
        pages = {304},
          doi = {10.3847/1538-3881/ae589a},
       adsurl = {https://ui.adsabs.harvard.edu/abs/2026AJ....171..304L}
}

@ARTICLE{Pontoppidan24,
       author = {{Pontoppidan}, Klaus M. and {Salyk}, Colette and {Banzatti}, Andrea and {Zhang}, Ke and {Pascucci}, Ilaria and {{\"O}berg}, Karin I. and {Long}, Feng and {Romero-Mirza}, Carlos E. and {Carr}, John and {Najita}, Joan and {Blake}, Geoffrey A. and {Arulanantham}, Nicole and {Andrews}, Sean and {Ballering}, Nicholas P. and {Bergin}, Edwin and {Calahan}, Jenny and {Cobb}, Douglas and {Colmenares}, Maria Jose and {Dickson-Vandervelde}, Annie and {Dignan}, Anna and {Green}, Joel and {Heretz}, Phoebe and {Herczeg}, Gregory and {Kalyaan}, Anusha and {Krijt}, Sebastiaan and {Pauly}, Tyler and {Pinilla}, Paola and {Trapman}, Leon and {Xie}, Chengyan},
        title = "{High-contrast JWST-MIRI Spectroscopy of Planet-forming Disks for the JDISC Survey}",
      journal = {\apj},
         year = 2024,
        month = mar,
       volume = {963},
       number = {2},
          eid = {158},
        pages = {158},
          doi = {10.3847/1538-4357/ad20f0},
archivePrefix = {arXiv},
       eprint = {2311.17020},
 primaryClass = {astro-ph.EP},
       adsurl = {https://ui.adsabs.harvard.edu/abs/2024ApJ...963..158P}
}

@Article{	  astropy:2013,
  adsurl	= {http://adsabs.harvard.edu/abs/2013A%26A...558A..33A},
  archiveprefix	= {arXiv},
  author	= {{Astropy Collaboration} and {Robitaille}, T.~P. and
		  {Tollerud}, E.~J. and {Greenfield}, P. and {Droettboom}, M.
		  and {Bray}, E. and {Aldcroft}, T. and {Davis}, M. and
		  {Ginsburg}, A. and {Price-Whelan}, A.~M. and {Kerzendorf},
		  W.~E. and {Conley}, A. and {Crighton}, N. and {Barbary}, K.
		  and {Muna}, D. and {Ferguson}, H. and {Grollier}, F. and
		  {Parikh}, M.~M. and {Nair}, P.~H. and {Unther}, H.~M. and
		  {Deil}, C. and {Woillez}, J. and {Conseil}, S. and
		  {Kramer}, R. and {Turner}, J.~E.~H. and {Singer}, L. and
		  {Fox}, R. and {Weaver}, B.~A. and {Zabalza}, V. and
		  {Edwards}, Z.~I. and {Azalee Bostroem}, K. and {Burke},
		  D.~J. and {Casey}, A.~R. and {Crawford}, S.~M. and
		  {Dencheva}, N. and {Ely}, J. and {Jenness}, T. and
		  {Labrie}, K. and {Lim}, P.~L. and {Pierfederici}, F. and
		  {Pontzen}, A. and {Ptak}, A. and {Refsdal}, B. and
		  {Servillat}, M. and {Streicher}, O.},
  doi		= {10.1051/0004-6361/201322068},
  eid		= {A33},
  eprint	= {1307.6212},
  journal	= {\aap},
  month		= oct,
  pages		= {A33},
  primaryclass	= {astro-ph.IM},
  title		= {{Astropy: A community Python package for astronomy}},
  volume	= 558,
  year		= 2013
}

@Article{	  astropy:2018,
  adsurl	= {https://ui.adsabs.harvard.edu/#abs/2018AJ....156..123T},
  author	= {{Price-Whelan}, A.~M. and {Sip{\H{o}}cz}, B.~M. and
		  {G{\"u}nther}, H.~M. and {Lim}, P.~L. and {Crawford}, S.~M.
		  and {Conseil}, S. and {Shupe}, D.~L. and {Craig}, M.~W. and
		  {Dencheva}, N. and {Ginsburg}, A. and {VanderPlas}, J.~T.
		  and {Bradley}, L.~D. and {P{\'e}rez-Su{\'a}rez}, D. and {de
		  Val-Borro}, M. and {Paper Contributors}, (Primary and
		  {Aldcroft}, T.~L. and {Cruz}, K.~L. and {Robitaille}, T.~P.
		  and {Tollerud}, E.~J. and {Coordination Committee},
		  (Astropy and {Ardelean}, C. and {Babej}, T. and {Bach},
		  Y.~P. and {Bachetti}, M. and {Bakanov}, A.~V. and
		  {Bamford}, S.~P. and {Barentsen}, G. and {Barmby}, P. and
		  {Baumbach}, A. and {Berry}, K.~L. and {Biscani}, F. and
		  {Boquien}, M. and {Bostroem}, K.~A. and {Bouma}, L.~G. and
		  {Brammer}, G.~B. and {Bray}, E.~M. and {Breytenbach}, H.
		  and {Buddelmeijer}, H. and {Burke}, D.~J. and {Calderone},
		  G. and {Cano Rodr{\'\i}guez}, J.~L. and {Cara}, M. and
		  {Cardoso}, J.~V.~M. and {Cheedella}, S. and {Copin}, Y. and
		  {Corrales}, L. and {Crichton}, D. and
		  {D{\textquoteright}Avella}, D. and {Deil}, C. and
		  {Depagne}, {\'E}. and {Dietrich}, J.~P. and {Donath}, A.
		  and {Droettboom}, M. and {Earl}, N. and {Erben}, T. and
		  {Fabbro}, S. and {Ferreira}, L.~A. and {Finethy}, T. and
		  {Fox}, R.~T. and {Garrison}, L.~H. and {Gibbons}, S.~L.~J.
		  and {Goldstein}, D.~A. and {Gommers}, R. and {Greco}, J.~P.
		  and {Greenfield}, P. and {Groener}, A.~M. and {Grollier},
		  F. and {Hagen}, A. and {Hirst}, P. and {Homeier}, D. and
		  {Horton}, A.~J. and {Hosseinzadeh}, G. and {Hu}, L. and
		  {Hunkeler}, J.~S. and {Ivezi{\'c}}, {\v{Z}}. and {Jain}, A.
		  and {Jenness}, T. and {Kanarek}, G. and {Kendrew}, S. and
		  {Kern}, N.~S. and {Kerzendorf}, W.~E. and {Khvalko}, A. and
		  {King}, J. and {Kirkby}, D. and {Kulkarni}, A.~M. and
		  {Kumar}, A. and {Lee}, A. and {Lenz}, D. and {Littlefair},
		  S.~P. and {Ma}, Z. and {Macleod}, D.~M. and {Mastropietro},
		  M. and {McCully}, C. and {Montagnac}, S. and {Morris},
		  B.~M. and {Mueller}, M. and {Mumford}, S.~J. and {Muna}, D.
		  and {Murphy}, N.~A. and {Nelson}, S. and {Nguyen}, G.~H.
		  and {Ninan}, J.~P. and {N{\"o}the}, M. and {Ogaz}, S. and
		  {Oh}, S. and {Parejko}, J.~K. and {Parley}, N. and
		  {Pascual}, S. and {Patil}, R. and {Patil}, A.~A. and
		  {Plunkett}, A.~L. and {Prochaska}, J.~X. and {Rastogi}, T.
		  and {Reddy Janga}, V. and {Sabater}, J. and {Sakurikar}, P.
		  and {Seifert}, M. and {Sherbert}, L.~E. and
		  {Sherwood-Taylor}, H. and {Shih}, A.~Y. and {Sick}, J. and
		  {Silbiger}, M.~T. and {Singanamalla}, S. and {Singer},
		  L.~P. and {Sladen}, P.~H. and {Sooley}, K.~A. and
		  {Sornarajah}, S. and {Streicher}, O. and {Teuben}, P. and
		  {Thomas}, S.~W. and {Tremblay}, G.~R. and {Turner},
		  J.~E.~H. and {Terr{\'o}n}, V. and {van Kerkwijk}, M.~H. and
		  {de la Vega}, A. and {Watkins}, L.~L. and {Weaver}, B.~A.
		  and {Whitmore}, J.~B. and {Woillez}, J. and {Zabalza}, V.
		  and {Contributors}, (Astropy},
  doi		= {10.3847/1538-3881/aabc4f},
  eid		= {123},
  journal	= {\aj},
  month		= sep,
  pages		= {123},
  primaryclass	= {astro-ph.IM},
  title		= {{The Astropy Project: Building an Open-science Project and
		  Status of the v2.0 Core Package}},
  volume	= {156},
  year		= 2018
}

@Article{	  astropy:2022,
  author	= {{Astropy Collaboration} and {Price-Whelan}, Adrian M. and
		  {Lim}, Pey Lian and {Earl}, Nicholas and {Starkman},
		  Nathaniel and {Bradley}, Larry and {Shupe}, David L. and
		  {Patil}, Aarya A. and {Corrales}, Lia and {Brasseur}, C.~E.
		  and {N{\"o}the}, Maximilian and {Donath}, Axel and
		  {Tollerud}, Erik and {Morris}, Brett M. and {Ginsburg},
		  Adam and {Vaher}, Eero and {Weaver}, Benjamin A. and
		  {Tocknell}, James and {Jamieson}, William and {van
		  Kerkwijk}, Marten H. and {Robitaille}, Thomas P. and
		  {Merry}, Bruce and {Bachetti}, Matteo and {G{\"u}nther}, H.
		  Moritz and {Aldcroft}, Thomas L. and {Alvarado-Montes},
		  Jaime A. and {Archibald}, Anne M. and {B{\'o}di}, Attila
		  and {Bapat}, Shreyas and {Barentsen}, Geert and
		  {Baz{\'a}n}, Juanjo and {Biswas}, Manish and {Boquien},
		  M{\'e}d{\'e}ric and {Burke}, D.~J. and {Cara}, Daria and
		  {Cara}, Mihai and {Conroy}, Kyle E. and {Conseil}, Simon
		  and {Craig}, Matthew W. and {Cross}, Robert M. and {Cruz},
		  Kelle L. and {D'Eugenio}, Francesco and {Dencheva}, Nadia
		  and {Devillepoix}, Hadrien A.~R. and {Dietrich}, J{\"o}rg
		  P. and {Eigenbrot}, Arthur Davis and {Erben}, Thomas and
		  {Ferreira}, Leonardo and {Foreman-Mackey}, Daniel and
		  {Fox}, Ryan and {Freij}, Nabil and {Garg}, Suyog and
		  {Geda}, Robel and {Glattly}, Lauren and {Gondhalekar}, Yash
		  and {Gordon}, Karl D. and {Grant}, David and {Greenfield},
		  Perry and {Groener}, Austen M. and {Guest}, Steve and
		  {Gurovich}, Sebastian and {Handberg}, Rasmus and {Hart},
		  Akeem and {Hatfield-Dodds}, Zac and {Homeier}, Derek and
		  {Hosseinzadeh}, Griffin and {Jenness}, Tim and {Jones},
		  Craig K. and {Joseph}, Prajwel and {Kalmbach}, J. Bryce and
		  {Karamehmetoglu}, Emir and {Ka{\l}uszy{\'n}ski}, Miko{\l}aj
		  and {Kelley}, Michael S.~P. and {Kern}, Nicholas and
		  {Kerzendorf}, Wolfgang E. and {Koch}, Eric W. and
		  {Kulumani}, Shankar and {Lee}, Antony and {Ly}, Chun and
		  {Ma}, Zhiyuan and {MacBride}, Conor and {Maljaars}, Jakob
		  M. and {Muna}, Demitri and {Murphy}, N.~A. and {Norman},
		  Henrik and {O'Steen}, Richard and {Oman}, Kyle A. and
		  {Pacifici}, Camilla and {Pascual}, Sergio and
		  {Pascual-Granado}, J. and {Patil}, Rohit R. and {Perren},
		  Gabriel I. and {Pickering}, Timothy E. and {Rastogi}, Tanuj
		  and {Roulston}, Benjamin R. and {Ryan}, Daniel F. and
		  {Rykoff}, Eli S. and {Sabater}, Jose and {Sakurikar},
		  Parikshit and {Salgado}, Jes{\'u}s and {Sanghi}, Aniket and
		  {Saunders}, Nicholas and {Savchenko}, Volodymyr and
		  {Schwardt}, Ludwig and {Seifert-Eckert}, Michael and
		  {Shih}, Albert Y. and {Jain}, Anany Shrey and {Shukla},
		  Gyanendra and {Sick}, Jonathan and {Simpson}, Chris and
		  {Singanamalla}, Sudheesh and {Singer}, Leo P. and
		  {Singhal}, Jaladh and {Sinha}, Manodeep and {Sip{\H{o}}cz},
		  Brigitta M. and {Spitler}, Lee R. and {Stansby}, David and
		  {Streicher}, Ole and {{\v{S}}umak}, Jani and {Swinbank},
		  John D. and {Taranu}, Dan S. and {Tewary}, Nikita and
		  {Tremblay}, Grant R. and {Val-Borro}, Miguel de and {Van
		  Kooten}, Samuel J. and {Vasovi{\'c}}, Zlatan and {Verma},
		  Shresth and {de Miranda Cardoso}, Jos{\'e} Vin{\'\i}cius
		  and {Williams}, Peter K.~G. and {Wilson}, Tom J. and
		  {Winkel}, Benjamin and {Wood-Vasey}, W.~M. and {Xue}, Rui
		  and {Yoachim}, Peter and {Zhang}, Chen and {Zonca}, Andrea
		  and {Astropy Project Contributors}},
  title		= "{The Astropy Project: Sustaining and Growing a
		  Community-oriented Open-source Project and the Latest Major
		  Release (v5.0) of the Core Package}",
  journal	= {\apj},
  year		= 2022,
  month		= aug,
  volume	= {935},
  number	= {2},
  eid		= {167},
  pages		= {167},
  doi		= {10.3847/1538-4357/ac7c74},
  archiveprefix	= {arXiv},
  eprint	= {2206.14220},
  primaryclass	= {astro-ph.IM},
  adsurl	= {https://ui.adsabs.harvard.edu/abs/2022ApJ...935..167A}
}

@ARTICLE{Gordon22,
       author = {{Gordon}, Karl D. and {Bohlin}, Ralph and {Sloan}, G.~C. and {Rieke}, George and {Volk}, Kevin and {Boyer}, Martha and {Muzerolle}, James and {Schlawin}, Everett and {Deustua}, Susana E. and {Hines}, Dean C. and et al.},
        title = "{The James Webb Space Telescope Absolute Flux Calibration. I. Program Design and Calibrator Stars}",
      journal = {\aj},
         year = 2022,
        month = jun,
       volume = {163},
       number = {6},
          eid = {267},
        pages = {267},
          doi = {10.3847/1538-3881/ac66dc},
archivePrefix = {arXiv},
       eprint = {2204.06500},
 primaryClass = {astro-ph.IM},
       adsurl = {https://ui.adsabs.harvard.edu/abs/2022AJ....163..267G}
}

@Article{	  massa20,
  author	= {{Massa}, Derck and {Fitzpatrick}, E.~L. and {Gordon}, Karl
		  D.},
  title		= "{An Analysis of the Shapes of Interstellar Extinction
		  Curves. VIII. The Optical Extinction Structure}",
  journal	= {\apj},
  year		= 2020,
  month		= mar,
  volume	= {891},
  number	= {1},
  eid		= {67},
  pages		= {67},
  doi		= {10.3847/1538-4357/ab6f01},
  archiveprefix	= {arXiv},
  eprint	= {2001.10880},
  primaryclass	= {astro-ph.SR},
  adsurl	= {https://ui.adsabs.harvard.edu/abs/2020ApJ...891...67M}
}

@ARTICLE{Pendleton25,
       author = {{Pendleton}, Yvonne J. and {Geballe}, T.~R. and {Chu}, Laurie E.~U. and {Decleir}, Marjorie and {Gordon}, Karl D. and {Tielens}, A.~G.~G.~M. and {Allamandola}, Louis J. and {Bouwman}, Jeroen and {Chiar}, J.~E. and {Dewitt}, Curtis and et al.},
        title = "{A Tale of Two Sightlines: Comparison of Hydrocarbon Dust Absorption Bands toward Cygnus OB2-12 and the Galactic Center}",
      journal = {\apj},
         year = 2025,
        month = oct,
       volume = {992},
       number = {1},
          eid = {8},
        pages = {8},
          doi = {10.3847/1538-4357/adfc3d},
archivePrefix = {arXiv},
       eprint = {2508.12601},
 primaryClass = {astro-ph.GA},
       adsurl = {https://ui.adsabs.harvard.edu/abs/2025ApJ...992....8P}
}

@Article{Hunter:2007,
  Author    = {Hunter, J. D.},
  Title     = {Matplotlib: A 2D graphics environment},
  Journal   = {Computing in Science \& Engineering},
  Volume    = {9},
  Number    = {3},
  Pages     = {90--95},
  publisher = {IEEE COMPUTER SOC},
  doi       = {10.1109/MCSE.2007.55},
  year      = 2007
}

@article{numpy,
 title         = {Array programming with {NumPy}},
 author        = {Charles R. Harris and K. Jarrod Millman and St{\'{e}}fan J.
                 van der Walt and Ralf Gommers and Pauli Virtanen and David
                 Cournapeau and Eric Wieser and Julian Taylor and Sebastian
                 Berg and Nathaniel J. Smith and Robert Kern and Matti Picus
                 and Stephan Hoyer and Marten H. van Kerkwijk and Matthew
                 Brett and Allan Haldane and Jaime Fern{\'{a}}ndez del
                 R{\'{i}}o and Mark Wiebe and Pearu Peterson and Pierre
                 G{\'{e}}rard-Marchant and Kevin Sheppard and Tyler Reddy and
                 Warren Weckesser and Hameer Abbasi and Christoph Gohlke and
                 Travis E. Oliphant},
 year          = {2020},
 month         = sep,
 journal       = {Nature},
 volume        = {585},
 number        = {7825},
 pages         = {357--362},
 doi           = {10.1038/s41586-020-2649-2},
 publisher     = {Springer Science and Business Media {LLC}},
 url           = {https://doi.org/10.1038/s41586-020-2649-2}
}

@book{python,
  author    = {Van Rossum, Guido and Drake, Fred L.},
  title     = {Python 3 Reference Manual},
  year      = {2009},
  isbn      = {1441412697},
  publisher = {CreateSpace},
  address   = {Scotts Valley, CA}
}

@software{scipy_18736568,
  author       = {Ralf Gommers and
                  Pauli Virtanen and
                  Matt Haberland and
                  Evgeni Burovski and
                  Tyler Reddy and
                  Warren Weckesser and
                  Travis E. Oliphant and
                  Andrew Nelson and
                  David Cournapeau and
                  Ilhan Polat and
                  alexbrc and
                  Pamphile Roy and
                  Pearu Peterson and
                  Lucas Colley and
                  Josh Wilson and
                  endolith and
                  Nikolay Mayorov and
                  Jake Bowhay and
                  Stefan van der Walt and
                  Albert Steppi and
                  Matthew Brett and
                  Denis Laxalde and
                  Eric Larson and
                  Atsushi Sakai and
                  Jarrod Millman and
                  Lars and
                  peterbell10 and
                  CJ Carey and
                  Paul van Mulbregt and
                  eric-jones},
  title        = {scipy/scipy: SciPy 1.17.1},
  month        = feb,
  year         = 2026,
  publisher    = {Zenodo},
  version      = {v1.17.1},
  doi          = {10.5281/zenodo.18736568},
  url          = {https://doi.org/10.5281/zenodo.18736568},
  swhid        = {swh:1:dir:934360229a7c597c39d811831bb6c020ef7f8151
                   ;origin=https://doi.org/10.5281/zenodo.595738;visi
                   t=swh:1:snp:2125cfcb4a989632c0e87df14bff9dc83bc816
                   e7;anchor=swh:1:rel:66e87815fdd0136976a21a8a0284e5
                   5c4e5b156e;path=scipy-scipy-c59ed93
                  },
}

@ARTICLE{2020SciPy-NMeth,
  author  = {Virtanen, Pauli and Gommers, Ralf and Oliphant, Travis E. and
            Haberland, Matt and Reddy, Tyler and Cournapeau, David and
            Burovski, Evgeni and Peterson, Pearu and Weckesser, Warren and
            Bright, Jonathan and {van der Walt}, St{\'e}fan J. and
            Brett, Matthew and Wilson, Joshua and Millman, K. Jarrod and
            Mayorov, Nikolay and Nelson, Andrew R. J. and Jones, Eric and
            Kern, Robert and Larson, Eric and Carey, C J and
            Polat, {\.I}lhan and Feng, Yu and Moore, Eric W. and
            {VanderPlas}, Jake and Laxalde, Denis and Perktold, Josef and
            Cimrman, Robert and Henriksen, Ian and Quintero, E. A. and
            Harris, Charles R. and Archibald, Anne M. and
            Ribeiro, Ant{\^o}nio H. and Pedregosa, Fabian and
            {van Mulbregt}, Paul and {SciPy 1.0 Contributors}},
  title   = {{{SciPy} 1.0: Fundamental Algorithms for Scientific
            Computing in Python}},
  journal = {Nature Methods},
  year    = {2020},
  volume  = {17},
  pages   = {261--272},
  adsurl  = {https://rdcu.be/b08Wh},
  doi     = {10.1038/s41592-019-0686-2},
}

@software{jwst_17515973,
  author       = {Bushouse, Howard and
                  Eisenhamer, Jonathan and
                  Dencheva, Nadia and
                  Davies, James and
                  Greenfield, Perry and
                  Morrison, Jane and
                  Hodge, Phil and
                  Simon, Bernie and
                  Grumm, David and
                  Droettboom, Michael and
                  Slavich, Edward and
                  Sosey, Megan and
                  Pauly, Tyler and
                  Miller, Todd and
                  Jedrzejewski, Robert and
                  Hack, Warren and
                  Davis, David and
                  Crawford, Steven and
                  Law, David and
                  Gordon, Karl and
                  Regan, Michael and
                  Cara, Mihai and
                  MacDonald, Ken and
                  Bradley, Larry and
                  Shanahan, Clare and
                  Jamieson, William and
                  Teodoro, Mairan and
                  Williams, Thomas and
                  Pena-Guerrero, Maria and
                  Graham, Brett and
                  Molter, Edward and
                  Brandt, Timothy and
                  Hayes, Christian and
                  Cooper, Rachel and
                  Clarke, Melanie and
                  Filippazzo, Joseph},
  title        = {JWST Calibration Pipeline},
  month        = nov,
  year         = 2025,
  publisher    = {Zenodo},
  version      = {1.20.2},
  doi          = {10.5281/zenodo.17515973},
  url          = {https://doi.org/10.5281/zenodo.17515973},
  swhid        = {swh:1:dir:fc8e0b17375bd6292e6f2fe6b758a3e44b81aa01
                   ;origin=https://doi.org/10.5281/zenodo.6984365;vis
                   it=swh:1:snp:ee0c72d562544a3226903308be9ab3969e139
                   858;anchor=swh:1:rel:4164a7a6a89b54a00ff5ed5b74c67
                   5e41f26b3fc;path=spacetelescope-jwst-ee52a96
                  },
}

@software{Photutils_17129028,
  author       = {Larry Bradley and
                  Brigitta Sipőcz and
                  Thomas Robitaille and
                  Erik Tollerud and
                  Zé Vinícius and
                  Christoph Deil and
                  Kyle Barbary and
                  Tom J Wilson and
                  Ivo Busko and
                  Axel Donath and
                  Hans Moritz Günther and
                  Mihai Cara and
                  P. L. Lim and
                  Sebastian Meßlinger and
                  Simon Conseil and
                  Michael Droettboom and
                  Azalee Bostroem and
                  E. M. Bray and
                  Lars Andersen Bratholm and
                  Zach Burnett and
                  William Jamieson and
                  Adam Ginsburg and
                  Dan Taranu and
                  Geert Barentsen and
                  Matt Craig and
                  Brett M. Morris and
                  Marshall Perrin and
                  Shivangee Rathi},
  title        = {astropy/photutils: 2.3.0},
  month        = sep,
  year         = 2025,
  publisher    = {Zenodo},
  version      = {2.3.0},
  doi          = {10.5281/zenodo.17129028},
  url          = {https://doi.org/10.5281/zenodo.17129028},
  swhid        = {swh:1:dir:dd51869167d76d722ba87e3f80f9f4199ec08c3f
                   ;origin=https://doi.org/10.5281/zenodo.596036;visi
                   t=swh:1:snp:30a5f50b0586911dc674668853d9abc352a2bc
                   22;anchor=swh:1:rel:e97861da904cf010c499a4211cd8a6
                   12373e912a;path=astropy-photutils-2294e35
                  },
}

@article{scikit-image,
 title = {scikit-image: image processing in {P}ython},
 author = {van der Walt, {S}t\'efan and {S}ch\"onberger, {J}ohannes {L}. and
           {Nunez-Iglesias}, {J}uan and {B}oulogne, {F}ran\c{c}ois and {W}arner,
           {J}oshua {D}. and {Y}ager, {N}eil and {G}ouillart, {E}mmanuelle and
           {Y}u, {T}ony and the scikit-image contributors},
 year = {2014},
 month = {6},
 volume = {2},
 pages = {e453},
 journal = {PeerJ},
 issn = {2167-8359},
 url = {https://doi.org/10.7717/peerj.453},
 doi = {10.7717/peerj.453}
}

@ARTICLE{software-citation-station-paper,
       author = {{Wagg}, Tom and {Broekgaarden}, Floor S.},
        title = "{Streamlining and standardizing software citations with The Software Citation Station}",
      journal = {arXiv e-prints},
         year = 2024,
        month = jun,
          eid = {arXiv:2406.04405},
        pages = {arXiv:2406.04405},
archivePrefix = {arXiv},
       eprint = {2406.04405},
 primaryClass = {astro-ph.IM},
       adsurl = {https://ui.adsabs.harvard.edu/abs/2024arXiv240604405W}
}

@software{software-citation-station-zenodo,
  author       = {Tom Wagg and
                  Floor Broekgaarden and
                  Phil Van-Lane and
                  Kai Wu and
                  Kayhan Gültekin},
  title        = {TomWagg/software-citation-station: v1.4},
  month        = nov,
  year         = 2025,
  publisher    = {Zenodo},
  version      = {v1.4},
  doi          = {10.5281/zenodo.17654855},
  url          = {https://doi.org/10.5281/zenodo.17654855},
  swhid        = {swh:1:dir:9a009430037c791424a572f542e9a5d5c1fb44ff
                   ;origin=https://doi.org/10.5281/zenodo.13225526;vi
                   sit=swh:1:snp:ef11f058d718d691f0661c9445c2251328cb
                   ac95;anchor=swh:1:rel:84cde4e532032537b7f014c4467c
                   102430a53fa2;path=TomWagg-software-citation-
                   station-61a588a
                  },
}

@software{mrspfpc,
  author       = {{Gordon}, Karl D. and {Law}, David R.},
  title        = {JWST MIRI Medium Resolution Spectrometer Point Fixed Pattern Corrections: Cleaner and Higher Signal-to-Noise 
  Spectra of Point Sources},
  month        = jul,
  year         = 2026,
  publisher    = {Zenodo},
  version      = {v1.0},
  doi          = {10.5281/zenodo.21263454},
  url          = {https://doi.org/10.5281/zenodo.21263454},
}
\bibliographystyle{aasjournal}

\end{document}